\pdfoutput=1
\documentclass[11pt]{article}
\usepackage[preprint]{acl}

\usepackage{times}
\usepackage{latexsym}

\usepackage[T1]{fontenc}

\usepackage[utf8]{inputenc}
\usepackage{microtype}
\usepackage{inconsolata}

\usepackage{amsmath,amsthm}
\usepackage{pifont}
\usepackage{amssymb,multirow,mathrsfs}
\usepackage{wasysym}
\usepackage[mathscr]{euscript}
\usepackage{placeins}
\usepackage{tabularx}
\usepackage{bbm}

\usepackage{booktabs}

\usepackage{afterpage}
\usepackage{float}
\usepackage{subcaption} 

\usepackage{fancyhdr}

\usepackage{xcolor}
\definecolor{darkgreen}{rgb}{0.0, 0.5, 0.0}

\usepackage{graphicx}
\usepackage{adjustbox}
\newcommand{\mySpade}{\adjustbox{valign=c}{\includegraphics[height=1.5em]{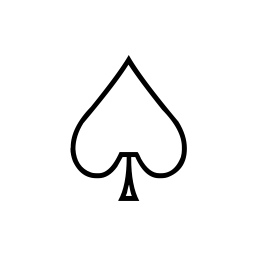}}}
\newcommand{\myClub}{\adjustbox{valign=c}{\includegraphics[height=1.5em]{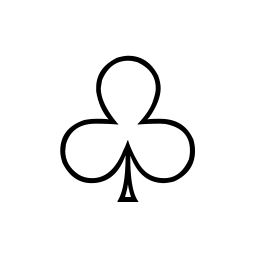}}}
\newcommand{\myDiamond}{\adjustbox{valign=c}{\includegraphics[height=1.5em]{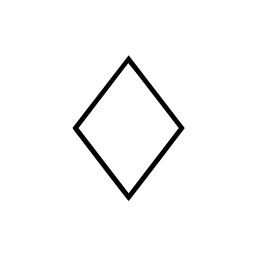}}}
\newcommand{\myHeart}{\adjustbox{valign=c}{\includegraphics[height=1.5em]{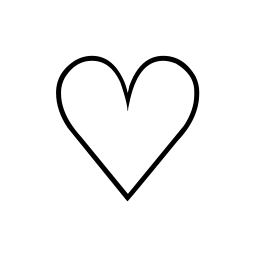}}}

\title{IR2: Information Regularization for Information Retrieval \thanks{Accepted by LREC-COLING 2024}}

\author{Jianyou Wang \thanks{Equal contribution} \quad
        Kaicheng Wang\footnotemark[2] \quad
        Xiaoyue Wang\footnotemark[2] \quad
        Weili Cao\\
        \textbf{Ramamohan Paturi \thanks{Equal Senior Authors} \quad
        Leon Bergen\footnotemark[3]} \\
        \\
        Laboratory for Emerging Intelligence\\
        University of California, San Diego \\
        \texttt{\{jiw101, kaw036, xiw027, w2cao, rpaturi, lbergen\}@ucsd.edu}
}
\begin{document}

\maketitle
\begin{abstract}
Effective information retrieval (IR) in settings with limited training data, particularly for complex queries, remains a challenging task. This paper introduces \textbf{IR2}, \textbf{I}nformation \textbf{R}egularization for \textbf{I}nformation \textbf{R}etrieval, a technique for reducing overfitting during synthetic data generation. This approach, representing a novel application of regularization techniques in synthetic data creation for IR, is tested on three recent IR tasks characterized by complex queries: DORIS-MAE, ArguAna, and WhatsThatBook. Experimental results indicate that our regularization techniques not only outperform previous synthetic query generation methods on the tasks considered but also reduce cost by up to 50\%. Furthermore, this paper categorizes and explores three regularization methods at different stages of the query synthesis pipeline—input, prompt, and output—each offering varying degrees of performance improvement compared to models where no regularization is applied. This provides a systematic approach for optimizing synthetic data generation in data-limited, complex-query IR scenarios. All code, prompts and synthetic data are available at \url{https://github.com/kwang927/Information-Regularization}.
\end{abstract}

\section{Introduction}
\label{sec: introduction}

Users often submit complex queries to information retrieval (IR) systems, expecting answers that accurately address multiple, interconnected questions. These complex queries may involve ambiguous language or multiple specific criteria, requiring that IR systems decipher and respond to the layered intents accurately to provide relevant results \cite{dorismae, whatsthatbook}. Ensuring that IR systems effectively handle the subtleties of such queries is important, allowing these systems to align with varied and nuanced user demands.

Effectively addressing complex queries in IR depends on a model's exposure to a wide range of user queries during training.
However, obtaining diverse real-world training data is often constrained by privacy concerns, availability, and resources. Synthetic data, therefore, becomes crucial, offering a means to expand training datasets, enabling models to learn from a broader spectrum of queries and user intents \cite{ma-etal-2021-zero, liang2020embedding}. 

\begin{figure*}
  \centering
  \includegraphics[width=\textwidth]{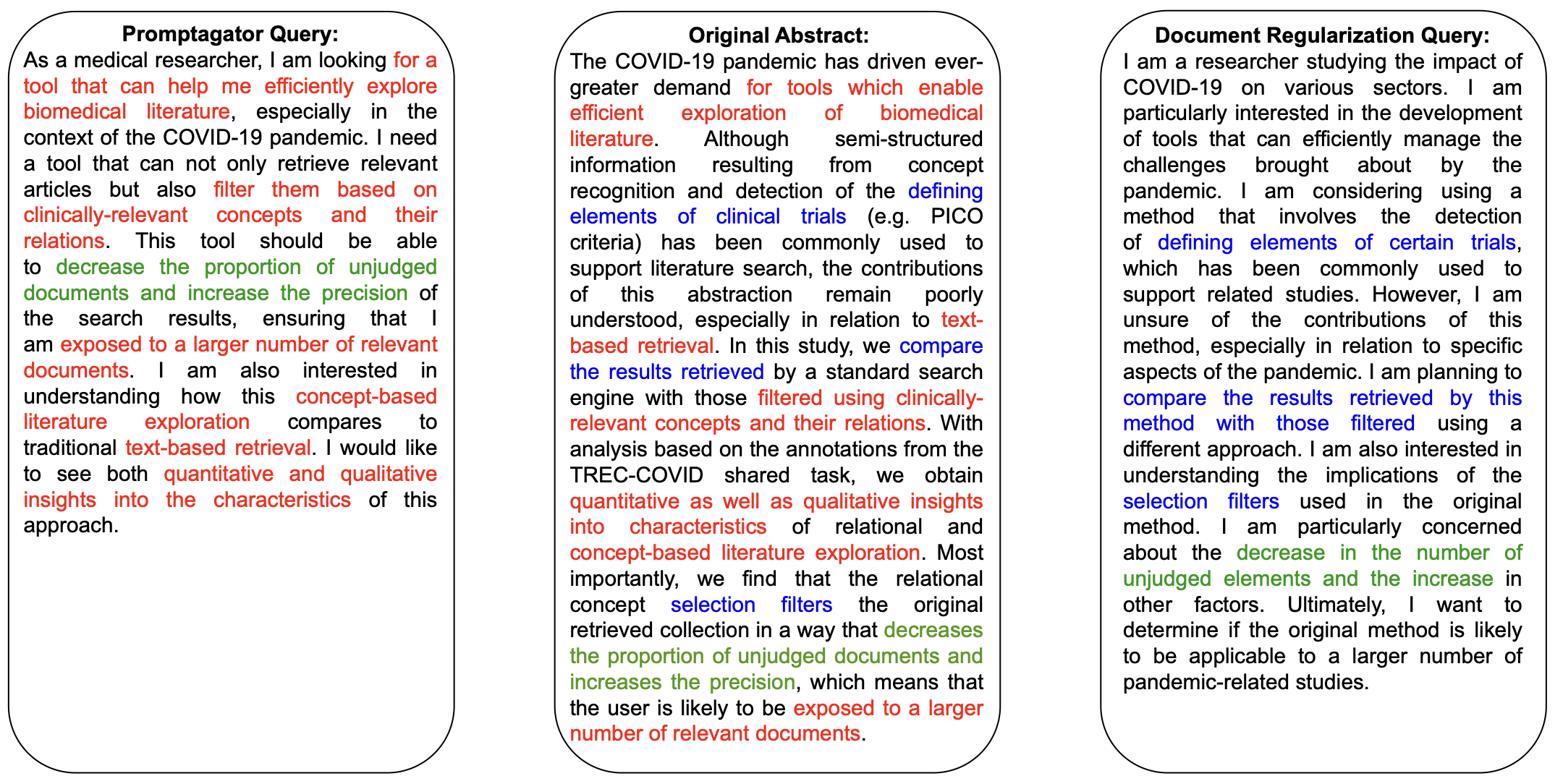}
  \caption{Sample synthetic query from Promptagator and a synthetic query generated with document regularization. (Both queries are generated from the same abstract.) \textcolor{red}{Red} indicates overlaps between the Promptagator query and the original abstract. \textcolor{blue}{Blue} indicates overlaps between the document regularized query and the original abstract. \textcolor{darkgreen}{Green} indicates overlap with both queries. The document regularized query has less textual overlap with the original query.}
  \label{fig:example query}
\end{figure*}

Recent research exploits Large Language Models (LLMs) to generate synthetic data pairs, constructing synthetic queries from real passages, often derived from zero-shot or few-shot examples \cite{bonifacio2022inpars, jeronymo2023inpars, meng2022augtriever, peng2023soft, controllable_QG}. 
Addressing the challenges of complex query information retrieval (IR) tasks through LLM-based synthetic data generation presents distinct difficulties. While synthetic data improves model performance across multiple tasks and metrics, generating queries from documents often results in synthetic pairs that exhibit superficial textual patterns, such as synonyms, matching keywords, and similar organizational flow, see the Promptagator  Query in Figure \ref{fig:example query} by a standard 8-shot synthetic query generation technique \cite{dai2022promptagator}. Though this might be effective for certain IR tasks, models may overfit to more superficial features, preventing them from understanding more conceptual relationships between query and document. 

\begin{figure}
  \centering
  \includegraphics[width=0.48\textwidth]{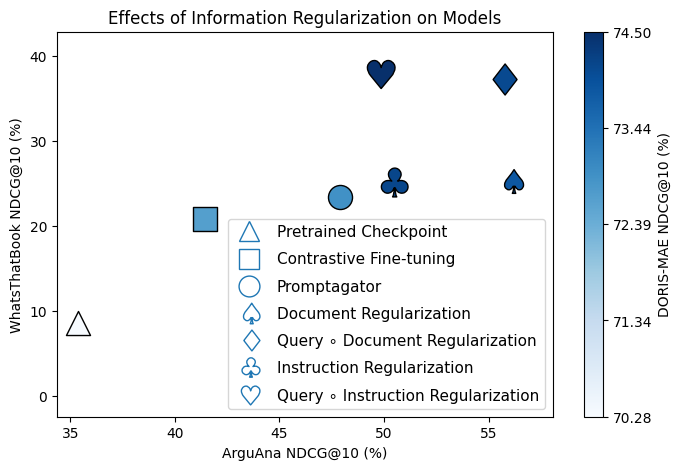}
  \caption{Performance of synthetic data generation methods on complex IR benchmarks. The $\triangle$, $\square$ \cite{gao-etal-2021-simcse}, and \textcircled{} \cite{dai2022promptagator} icons represent baselines. The other four icons \protect\mySpade \protect\myHeart \protect\myClub \protect\myDiamond denote IR2 approaches, indicating the performance of models after fine-tuning on information-regularized synthetic datasets. Metrics are chosen based on standard practice for the three benchmarks. Model performances are averaged across all models used in experiments. 
 }
  \label{fig:effects}
\end{figure}


This paper focuses on developing methods to generate and utilize synthetic data to enhance the retrieval capabilities of IR systems. To mitigate overfitting, we introduce IR2 (Information Regularization for Information Retrieval), which creates queries that have conceptual overlap with the original document, but varied phrasing and structure. This yields synthetic queries that have a less explicit relationship with the document, see the Document Regularization Query in Figure \ref{fig:example query}. We posit that training models with such data can be advantageous by nudging them towards understanding deeper semantic relationships, resulting in improved retrieval performance for complex queries.

In this paper, we make several contributions to synthetic data generation for Information Retrieval. First, we introduce three distinct regularization techniques applied at various stages of the synthetic data generation process. This approach addresses models' tendency to learn superficial and unnatural overlapping features between synthetic queries and documents, which would hamper model performance on complex query IR tasks

We also present a comprehensive analysis of how different pretrained transformer-based embedding models respond to these regularization techniques \cite{e5-large-v2, roberta, specter}. Our empirical evidence demonstrates consistent performance improvements with our methods compared to non-regularized baselines \cite{gao-etal-2021-simcse, dai2022promptagator}.

Furthermore, we find that these IR2 regularization techniques can be effectively combined, with instruction and output regularization emerging as the most potent pairing in our tests, as shown by \myHeart and \myDiamond in Figure \ref{fig:effects}, top right corner. This finding paves the way for more effective strategies in data augmentation for IR systems.

\section{Related Work}
\label{sec: related works}

In recent years, Information Retrieval has seen increasing task complexity as well as advancements in models designed to navigate this greater complexity.

IR tasks have typically centered on retrieving relevant passages to answer simple, sentence-level queries, as seen in foundational datasets like MS MARCO \cite{msmarco} and NQ \cite{NQ}. The BEIR dataset \cite{thakur2021beir} combines 19 IR tasks, most of which are sentence-level, including specific challenges like TREC-COVID \cite{treccovid} and SCIFACT \cite{scifact}. Several recent datasets contain complex, paragraph-length queries. DORIS-MAE \cite{dorismae} poses the task of retrieving scientific abstracts given complex scientific research questions. In the WhatsThatBook dataset \cite{whatsthatbook}, tip-of-the-tongue user queries are used to retrieve book descriptions. In ArguAna \cite{arguana}, paragraph-length arguments are used to retrieve counterarguments.

Lexicon-based methods such as TF-IDF \cite{tfidf} and BM25 \cite{bm25} retrieve based on keyword matching, and are effective in scenarios with substantial token overlap between queries and relevant passages. However, their limitations became apparent as tasks required deeper semantic comprehension, especially in cases with minimal token overlap.

Transformer-based models, including various cross-encoders and dual-encoders \cite{e5-large-v2, simlm, gao-etal-2021-simcse, specter, colbertv2, ance, spladev2}, have offered more nuanced document and query representations. They have shown strong results on benchmarks like BEIR \cite{thakur2021beir} but have struggled on complex query tasks.

Recently, improvements in LLMs, coupled with prompting techniques such as Chain-of-Thought \cite{cot1, cot2} and In-Context-Learning \cite{icl1, icl3}, have influenced the field of Information Retrieval (IR) in several directions. This has included data annotation \cite{dorismae, gilardi2023chatgpt, wang-etal-2022-gpl} as well as retrieval and reranking processes \cite{sun2023chatgpt}. Most relevant for the current work, LLMs have been used for synthetic data generation, particularly in data-scarce conditions, outperforming traditional augmentation methods \cite{izacard2022unsupervised}. For instance, \citet{ma-etal-2021-zero, liang2020embedding} applied LLMs for synthetic question generation in Question Answering (QA). 

In IR tasks, since user-generated queries are frequently difficult or expensive to obtain, much of the work on synthetic data has focused on generating synthetic queries. This has included InPars \cite{bonifacio2022inpars, jeronymo2023inpars} and Promptagator \cite{dai2022promptagator}, the latter showcasing significant success on the BEIR benchmark. Augtriever \cite{meng2022augtriever} introduced methods for synthetic query generation using smaller models, optimizing both time and cost. \citet{peng2023soft} used soft prompt-tuning to further enhance the quality of generated queries.

Synthetic data generation also has applications in other fields, including text classification \cite{SDG_classification}, information extraction \cite{SDG_IE}, reinforcement learning for language model alignment \cite{SDG_RL}, human-computer interaction \cite{SDG_HCI}, and computational social science \cite{SDG_CSS}.

In addition to these LLM developments, there has been an ongoing push to improve sentence embeddings, including SimCSE \cite{gao-etal-2021-simcse}, which applies dropout masking as a data augmentation technique, and other variants such as DiffCSE \cite{chuang-etal-2022-diffcse}, and RankCSE \cite{liu-etal-2023-rankcse}. By using a contrastive learning objective, these models have substantially increased the utility of unlabeled data for retrieval systems.

\section{Methodology}
\label{sec: methodologies}

\begin{figure*}[ht]
    \centering

    \begin{subfigure}{0.3\textwidth}
        \includegraphics[width=0.9\linewidth]{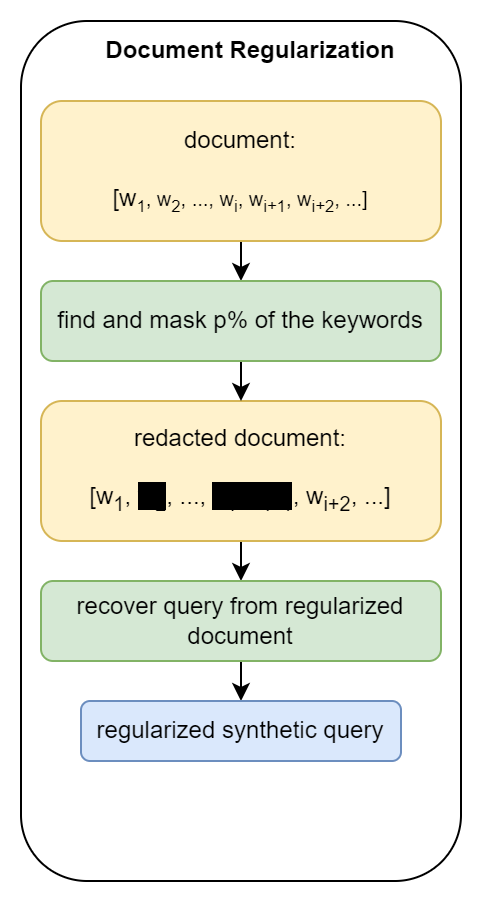}
        \caption{Document Regularization}
        \label{fig:document_regularization}
    \end{subfigure}
    \hfill
    \begin{subfigure}{0.3\textwidth}
        \includegraphics[width=0.9\linewidth]{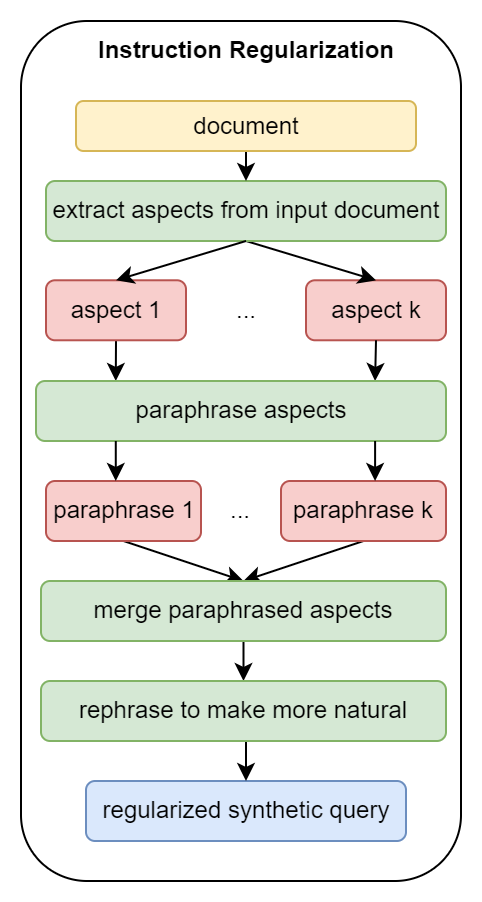}
        \caption{Instruction Regularization}
        \label{fig:instruction_regularization}
    \end{subfigure}
    \hfill
    \begin{subfigure}{0.3\textwidth}
        \includegraphics[width=0.9\linewidth]{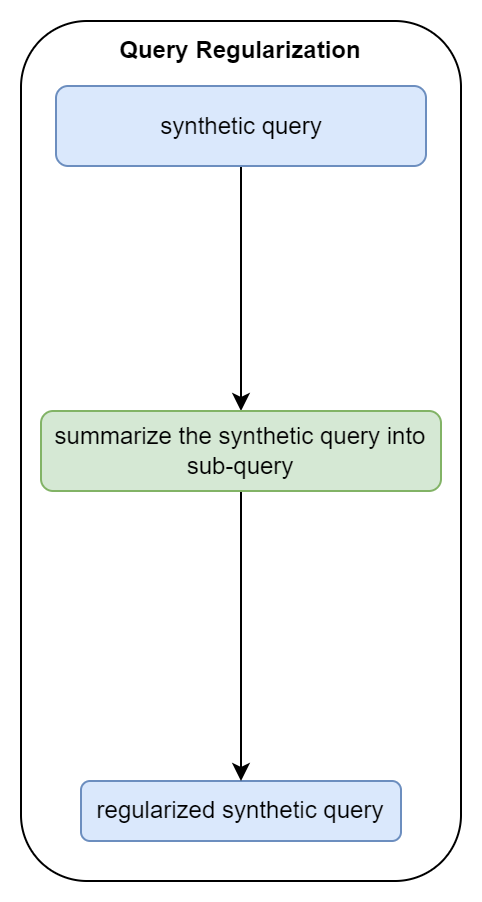}
        \caption{Query Regularization}
        \label{fig:query_regularization}
    \end{subfigure}

    \caption{Illustration of the three information regularization methods.}
    \label{fig:sigrid_pipline}
\end{figure*}

\subsection{Synthetic Query Generation}


The goal of IR systems is to match user queries, which express the user's information need, with relevant documents that contain the desired information. A typical IR system would benefit from fine-tuning of pairs of query and relevant document if sufficient data is available. However, real user queries are difficult and costly to collect. In addition, finding relevant documents for user queries also requires careful annotations. Therefore, synthetic query generation aims to improve IR system performance when faced with challenges of the scarcity of query data and the lack of supervised pairs of query and relevant document. Since documents are generally more readily available, by generating artificial yet plausible user queries based on existing documents, we can create additional training data for IR systems, helping them better understand and respond to a wider range of potential user queries.

In this research, we concentrate on methods for generating synthetic queries that are relevant to given documents. This involves not just creating queries that a user might realistically pose, but also ensuring that these queries are diverse enough to train IR systems effectively. By using this synthetic data for training, we aim to improve the systems' overall ability to handle complex user queries, enhancing retrieval accuracy and efficiency.

\subsection{Baseline Query Generation}
 Promptagator \cite{dai2022promptagator} proposes a few-shot method for LLM synthetic query generation. By providing an LLM with a small number of pairs of relevant documents and queries, it will implicitly learn the transformation from document to query. When a new document is provided, the LLM will transform it to a relevant query. We use $d_i$ to denote a document and $q_i$ to denote a relevant query. Promptagator uses 8-shot prompting of the following form:
$$d_1,q_1,\dots, d_8, q_8,d_9, $$
which includes the 8 example pairs as well as the target document $d_9$.

However, as shown in Figure \ref{fig:example query}, biases from the LLM may result in suboptimal queries. In our testing, we observed that queries frequently contained an implausibly large number of details from the documents, including highly similar phrasing and mirroring structure. When these queries are used for training downstream IR systems, they may lead to overfitting, as the IR systems learn simple phrase-matching heuristics between queries and documents.

\subsection{Information Regularization}

In order to mitigate these potential problems in the query generation process, we introduce IR2 (Information Regularization for Information Retrieval). These methods aim to reduce the number of superficial features shared by documents and synthetic queries, while still maintaining deeper semantic relationships.

The three information regularization methods can be categorized by the stage of the query generation pipeline at which they are applied: 
\begin{itemize}
    \item Document Regularization is applied to the input document $d_i$ of LLM by intentionally withholding parts of the semantic information in $d_i$.
    \item Instruction Regularization is applied to the prompt. We design a specialized prompt for the LLM, explicitly instructing it to avoid superficial similarities between the query and document.
    \item Query Regularization is applied to the synthetic query which was generated by the LLM. It is designed to summarize the synthetic query, transforming it into a less complex sub-query. 
\end{itemize}


Below we will discuss the motivation and intended effects of each type of regularization.

\subsubsection{Input Document Regularization}

\label{ssec: document regularization}

Document Regularization aims to address implausibly high levels of phrase overlaps between documents and synthetic queries. This technique disrupts phrase overlaps by partially masking the document's content before feeding it into the LLM. As shown in Figure \ref{fig:sigrid_pipline} (a), the system identifies key semantic words in a document and hides a random $p\%$ of them. The LLM, guided by the style of a single example query, then generates a new query based on this incomplete information.

This approach forces the LLM to generate queries that diverge textually from the source document while still remaining broadly relevant, and it makes the resulting training data more challenging for downstream IR systems. The method can be straightforwardly adapted to different IR datasets, and the $p\%$ parameter offers control over the degree of regularization. Setting $p = 0$ defaults to the original Promptagator approach, with no regularization.

\subsubsection{Instruction Regularization}
\label{ssec: prompt regularization}


\vspace{0.2cm}

This method involves guiding the LLM through a structured thought process to generate queries from documents, emphasizing the extraction and synthesis of key ideas while avoiding superficial mimicry. The model is first instructed to break the document into a sequence of aspects, each describing an important part of the document. Each aspect is subsequently paraphrased. The LLM is finally instructed to combine the paraphrased aspects into a natural query. The approach is detailed in Figure \ref{fig:sigrid_pipline} (b).

Due to the complexity of the prompt, we found that GPT-3.5 was unable to perform this task, while GPT-4 (which is used for our experiments) was able to.


\subsubsection{Output Query Regularization}
\label{ssec: query regularization}

As depicted in Figure \ref{fig:sigrid_pipline} (c), this technique simplifies synthetic queries, making them shorter and less detailed yet still conceptually relevant to the source document. The process challenges embedding models to focus on deeper semantic understanding rather than textual similarity.


\vspace{0.2cm}

Query regularization is straightforward, requires a small context window, and can be combined with our other synthetic generation techniques. In our experiments, we investigate the effectiveness of query regularization applied to synthetic queries generated from document regularization and instruction regularization. 


\section{Experiments}
\setlength{\tabcolsep}{3pt}

\label{sec: experiments}
To evaluate the effectiveness of information regularization for synthetic query generation, we choose 3 complex-query based test benchmarks (DORIS-MAE, ArguAna and WhatsThatBook) and 4 different embedding models (E5-v2-Large \cite{e5-large-v2}, SimCSE-Large \cite{gao-etal-2021-simcse}, RoBERTa-Large \cite{roberta}, SPECTER-v2 \cite{specter}). For each benchmark, we generate 8 synthetic datasets:
\begin{itemize}
    \item Document Regularization ($p=40\%$)
    \item Document Regularization ($p=60\%$)
    \item Document Regularization ($p=80\%$)
    \item Instruction Regularization
    \item Query Regularization $\circ$ Document Regularization ($p=40\%$) 
    \item Query Regularization $\circ$ Document Regularization ($p=60\%$) 
    \item Query Regularization $\circ$  Document Regularization ($p=80\%$) 
    \item Query Regularization $\circ$ Instruction Regularization. 
\end{itemize}
The notation $\circ$ indicates function composition.

In this section, we will discuss the details of benchmark datasets, pretrained models, baseline query generation methods, and regularization techniques. All experiment code is publicly accessible.\footnote{\url{https://github.com/Info-Regularization/Information-Regularization}}

For model evaluations, we report the IR metrics recommended by each benchmark's associated paper. Full experimental results with a more complete set of metrics are reported in Appendix \ref{sec: appendix_experiment_details}. 

\subsection{Benchmarks}
\label{ssec: datasets}

\textbf{WhatsThatBook (WTB)} \cite{whatsthatbook} contains 14,441 queries from users trying to recall specific books, paired with book titles and descriptions. The task is to match each of the 1,445 test set queries with the correct descriptions. We use descriptions from 4,000 unique books, exclusive from the test set, for synthetic query generation.


\textbf{DORIS-MAE} \cite{dorismae} includes 100 complex research queries in AI, CV, NLP, and ML, split into 40 for training and 60 for testing. Each query is associated with a candidate pool of approximately 100 research paper abstracts, with a fine-grained ranking system. The candidate pools are drawn from a corpus of approximately 360,000 computer science papers.
We use 4,000 abstracts from a 360,000 paper corpus for synthetic query generation.


\textbf{ArguAna} \cite{arguana} from BEIR \cite{thakur2021beir} consists of 8,674 paragraph-length arguments, and a test set of 1,406 argument-counterargument pairs. The task is to retrieve the correct counterargument for each test set argument. We use 5,700 arguments, not part of the test set, for synthetic counterargument generation. Due to the symmetry between arguments and counterarguments, we can use our synthetic data to train for retrieval of counterarguments given arguments.

\subsection{Models}
\label{ssec: models}
 
We experiment with four models: E5-Large-v2 \cite{e5-large-v2} (355M), SimCSE-Large \cite{gao-etal-2021-simcse} (355M), RoBERTa-Large \cite{roberta} (355M), and SPECTER-v2 \cite{specter} (110M). E5-Large-v2 (denoted as E5), a contrastively trained model, demonstrates performance on par with the state-of-the-art ada-002 \cite{openaiada} on BEIR \cite{thakur2021beir}. For SimCSE-Large (denoted as SimCSE), which uses self-supervised training with dropout masks, we choose the top-performing checkpoint on the STS benchmark \cite{sts}. RoBERTa-Large is denoted as RoBERTa. SPECTER-v2 is specialized for scientific document representation. Embeddings are derived according to each model's specifications, using pooling strategies, and retrieval and reranking are based on similarity measures (L2 distance for SPECTER-v2; cosine similarity for others).

\subsection{Baselines}
\label{ssec:baselines}
Three baseline types were used. First, the "Pretrained" baselines indicate performance of the pretrained checkpoints on each dataset. Second, a "Contrastive Fine-tuning" approach, similar to SimCSE \cite{gao-etal-2021-simcse}, applies random dropout masks to documents, creating self-supervised pairs for fine-tuning all models. This is used as a domain adaptation baseline. Third, we compare to Promptagator \cite{dai2022promptagator}, a few-shot synthetic query generation method. Its synthetic data generation process was replicated, using GPT-4 (gpt-4-0613) instead of FLAN 137B \cite{flan}. 

To eliminate possible data confounds, all baselines use the same data as the regularization methods. For each benchmark, the regularization methods use a subset of documents for synthetic query generation. For the Promptagator baseline, the same documents are used for synthetic generation. For the Contrastive Fine-tuning baseline, the embedding models are directly fine-tuned on the same documents. Specifically, we train using 4000 documents from DORIS-MAE’s corpus, 5700 arguments from ArguAna, and 4000 book descriptions from WhatsThatBook. We removed documents which were unusually short or long, and randomly sampled from the remaining documents.

\begin{table*}[!htbp]
\setlength{\tabcolsep}{1.5pt}
\centering
\scriptsize
\begin{tabular}{p{2.5cm}p{1.55cm}p{1.55cm}p{1.55cm}p{1.55cm}|p{1.55cm}p{1.55cm}p{1.55cm}p{1.55cm}}

\toprule
& \multicolumn{4}{c}{E5-Large-v2} & \multicolumn{4}{c}{RoBERTa} \\
\cmidrule(lr){2-5} \cmidrule(lr){6-9}
Method & R@5 & R@20 & NDCG@10 & MRR@10  & R@5 & R@20 & NDCG@10 & MRR@10 \\
\midrule
Pretrained & 18.41 & 27.06 & 15.25 & 13.07 & 2.84 & 5.33 & 2.17 & 1.73 \\
Contrastive Fine-tuning   & 37.25 & 49.85 & 33.16 & 29.87 & 23.59 & 35.27 & 20.69 & 18.10  \\
\cmidrule(lr){1-1}
Promptagator & 41.53 & 54.66 & 37.07 & 33.56 & 27.23 & 39.14 & 23.69 & 20.78 \\

\cmidrule(lr){1-1}

Doc\textsubscript{40\% reg} & 41.40 & 53.69 \color{red} $\downarrow$ -0.97  & 37.01 & 33.72 & 28.30 \color{darkgreen} $\uparrow$ 1.07  & 40.89 \color{darkgreen} $\uparrow$ 1.75  & 24.79 \color{darkgreen} $\uparrow$ 1.10  & 21.81 \color{darkgreen} $\uparrow$ 1.03  \\
Query\textsubscript{reg} $\circ$  Doc\textsubscript{40\% reg} & 45.50 \color{darkgreen} $\uparrow$ 3.97  & 58.08 \color{darkgreen} $\uparrow$ 3.42  & 40.50 \color{darkgreen} $\uparrow$ 3.43  & 37.01 \color{darkgreen} $\uparrow$ 3.45 & 32.83 \color{darkgreen} $\uparrow$ 5.60  & 46.38 \color{darkgreen} $\uparrow$ 7.24  & 29.05 \color{darkgreen} $\uparrow$ 5.36  & 25.82 \color{darkgreen} $\uparrow$ 5.04 \\

Doc\textsubscript{60\% reg}  & 44.15 \color{darkgreen} $\uparrow$ 2.62  & 55.69 \color{darkgreen} $\uparrow$ 1.03  & 39.64 \color{darkgreen} $\uparrow$ 2.57  & 36.16 \color{darkgreen} $\uparrow$ 2.60 & 28.67 \color{darkgreen} $\uparrow$ 1.44  & 41.49 \color{darkgreen} $\uparrow$ 2.35  & 25.14 \color{darkgreen} $\uparrow$ 1.45  & 22.06 \color{darkgreen} $\uparrow$ 1.28   \\

Query\textsubscript{reg} $\circ$  Doc\textsubscript{60\% reg} &  \textbf{46.23} \color{darkgreen} $\uparrow$ 4.70  & \textbf{59.76} \color{darkgreen} $\uparrow$ 5.10  & \textbf{41.58} \color{darkgreen} $\uparrow$ 4.51  & \textbf{38.13} \color{darkgreen} $\uparrow$ 4.57 & 34.30 \color{darkgreen} $\uparrow$ 7.07  & \textbf{47.52} \color{darkgreen} $\uparrow$ 8.38  & \textbf{30.15} \color{darkgreen} $\uparrow$ 6.46  & \textbf{26.83} \color{darkgreen} $\uparrow$ 6.05   \\

Doc\textsubscript{80\% reg} & 28.30 \color{red} $\downarrow$ -13.23  & 39.85 \color{red} $\downarrow$ -14.81  & 24.05 \color{red} $\downarrow$ -13.02  & 21.13 \color{red} $\downarrow$ -12.43 & 21.74 \color{red} $\downarrow$ -5.49  & 33.16 \color{red} $\downarrow$ -5.98  & 18.91 \color{red} $\downarrow$ -4.78  & 16.35 \color{red} $\downarrow$ -4.43  \\

Query\textsubscript{reg} $\circ$  Doc\textsubscript{80\% reg} & 44.80 \color{darkgreen} $\uparrow$ 3.27  & 58.21 \color{darkgreen} $\uparrow$ 3.55  & 40.29 \color{darkgreen} $\uparrow$ 3.22  & 36.94 \color{darkgreen} $\uparrow$ 3.38 & \textbf{33.69} \color{darkgreen} $\uparrow$ 6.46  & 46.23 \color{darkgreen} $\uparrow$ 7.09  & 29.41 \color{darkgreen} $\uparrow$ 5.72  & 26.20 \color{darkgreen} $\uparrow$ 5.42 \\

\cmidrule(lr){1-1}

Instr\textsubscript{reg} &  42.15 \color{darkgreen} $\uparrow$ 0.62  & 53.02 \color{red} $\downarrow$ -1.64  & 37.33 & 33.89 \color{darkgreen} $\uparrow$ 0.33 & 28.12 \color{darkgreen} $\uparrow$ 0.89  & 40.80 \color{darkgreen} $\uparrow$ 1.66  & 24.51 \color{darkgreen} $\uparrow$ 0.82  & 21.46 \color{darkgreen} $\uparrow$ 0.68 \\

Query\textsubscript{reg} $\circ$  Instr\textsubscript{reg} &  44.90 \color{darkgreen} $\uparrow$ 3.37  & 56.28 \color{darkgreen} $\uparrow$ 1.62  & 39.29 \color{darkgreen} $\uparrow$ 2.22  & 35.79 \color{darkgreen} $\uparrow$ 2.23 & 33.09 \color{darkgreen} $\uparrow$ 5.86  & 46.09 \color{darkgreen} $\uparrow$ 6.95  & 29.28 \color{darkgreen} $\uparrow$ 5.59  & 26.06 \color{darkgreen} $\uparrow$ 5.28  \\
\midrule
& \multicolumn{4}{c}{SimCSE} & \multicolumn{4}{c}{SPECTER-v2} \\
\cmidrule(lr){2-5} \cmidrule(lr){6-9}
Method & R@5 & R@20 & NDCG@10 & MRR@10  & R@5 & R@20 & NDCG@10 & MRR@10 \\
\midrule
Pretrained & 16.96 & 24.98 & 14.58 & 12.67 & 2.98 & 5.26 & 2.43 & 1.87 \\
Contrastive Fine-tuning & 26.78 & 37.15 & 23.11 & 20.42 & 7.85 & 12.56 & 6.36 & 5.35\\
\cmidrule(lr){1-1}
Promptagator & 30.97 & 44.26 & 27.15 & 23.99 & 6.56 & 11.81 & 5.72 & 4.88 \\
\cmidrule(lr){1-1}

Doc\textsubscript{40\% reg} & 32.10 \color{darkgreen} $\uparrow$ 1.13  & 43.07 \color{red} $\downarrow$ -1.19  & 28.30 \color{darkgreen} $\uparrow$ 1.15  & 25.45 \color{darkgreen} $\uparrow$ 1.46 & 9.98 \color{darkgreen} $\uparrow$ 3.42  & 16.12 \color{darkgreen} $\uparrow$ 4.31  & 8.55 \color{darkgreen} $\uparrow$ 2.83  & 7.32 \color{darkgreen} $\uparrow$ 2.44 \\

Query\textsubscript{reg} $\circ$ Doc\textsubscript{40\% reg} & 34.63 \color{darkgreen} $\uparrow$ 3.66  & 47.44 \color{darkgreen} $\uparrow$ 3.18  & 30.64 \color{darkgreen} $\uparrow$ 3.49  & 27.51 \color{darkgreen} $\uparrow$ 3.52 & 10.84 \color{darkgreen} $\uparrow$ 4.28  & 16.26 \color{darkgreen} $\uparrow$ 4.45  & 8.81 \color{darkgreen} $\uparrow$ 3.09  & 7.57 \color{darkgreen} $\uparrow$ 2.69  \\

Doc\textsubscript{60\% reg}  & 33.03 \color{darkgreen} $\uparrow$ 2.06  & 43.96 & 28.87 \color{darkgreen} $\uparrow$ 1.72  & 25.97 \color{darkgreen} $\uparrow$ 1.98 & 8.64 \color{darkgreen} $\uparrow$ 2.08  & 14.49 \color{darkgreen} $\uparrow$ 2.68  & 7.46 \color{darkgreen} $\uparrow$ 1.74  & 6.26 \color{darkgreen} $\uparrow$ 1.38   \\

Query\textsubscript{reg} $\circ$  Doc\textsubscript{60\% reg}  & 34.84 \color{darkgreen} $\uparrow$ 3.87  & 47.08 \color{darkgreen} $\uparrow$ 2.82  & 30.77 \color{darkgreen} $\uparrow$ 3.62  & 27.64 \color{darkgreen} $\uparrow$ 3.65 & 7.54 \color{darkgreen} $\uparrow$ 0.98  & 13.29 \color{darkgreen} $\uparrow$ 1.48  & 6.55 \color{darkgreen} $\uparrow$ 0.83  & 5.48 \color{darkgreen} $\uparrow$ 0.60  \\

Doc\textsubscript{80\% reg} & 19.75 \color{red} $\downarrow$ -11.22  & 29.89 \color{red} $\downarrow$ -14.37  & 16.73 \color{red} $\downarrow$ -10.42  & 14.33 \color{red} $\downarrow$ -9.66  & 7.39 \color{darkgreen} $\uparrow$ 0.83  & 13.19 \color{darkgreen} $\uparrow$ 1.38  & 6.46 \color{darkgreen} $\uparrow$ 0.74  & 5.31 \color{darkgreen} $\uparrow$ 0.43 \\

Query\textsubscript{reg} $\circ$  Doc\textsubscript{80\% reg} & 33.76 \color{darkgreen} $\uparrow$ 2.79  & 45.63 \color{darkgreen} $\uparrow$ 1.37  & 29.34 \color{darkgreen} $\uparrow$ 2.19  & 26.26 \color{darkgreen} $\uparrow$ 2.27 & 8.14 \color{darkgreen} $\uparrow$ 1.58  & 14.69 \color{darkgreen} $\uparrow$ 2.88  & 7.33 \color{darkgreen} $\uparrow$ 1.61  & 6.13 \color{darkgreen} $\uparrow$ 1.25 \\

\cmidrule(lr){1-1}

Instr\textsubscript{reg} & 33.81 \color{darkgreen} $\uparrow$ 2.84  & 45.79 \color{darkgreen} $\uparrow$ 1.53  & 29.74 \color{darkgreen} $\uparrow$ 2.59  & 26.79 \color{darkgreen} $\uparrow$ 2.80  & 10.49 \color{darkgreen} $\uparrow$ 3.93  & 16.80 \color{darkgreen} $\uparrow$ 4.99  & 9.30 \color{darkgreen} $\uparrow$ 3.58  & 8.07 \color{darkgreen} $\uparrow$ 3.19 \\

Query\textsubscript{reg} $\circ$ Instr\textsubscript{reg} & \textbf{37.34} \color{darkgreen} $\uparrow$ 6.37  & \textbf{50.45} \color{darkgreen} $\uparrow$ 6.19  & \textbf{33.18} \color{darkgreen} $\uparrow$ 6.03  & \textbf{30.02} \color{darkgreen} $\uparrow$ 6.03  & \textbf{11.08} \color{darkgreen} $\uparrow$ 4.52  & \textbf{17.36} \color{darkgreen} $\uparrow$ 5.55  & \textbf{9.59} \color{darkgreen} $\uparrow$ 3.87  & \textbf{8.28} \color{darkgreen} $\uparrow$ 3.40  \\

\bottomrule
\end{tabular}
\caption{Results for WhatsThatBook. The average of 20 random trials is reported for each model/method/metric. \textcolor{darkgreen}{Green arrow} indicates a statistically significant ($p<0.05$) increase over Promptagator baseline. \textcolor{red}{Red arrow} indicates a significant decrease.}
\label{tab: wtb table}
\end{table*}

\subsection{Implementation of Regularization}
\label{ssec: implementation of regularization}

\textbf{Input Document Regularization}: We use gpt-3.5-turbo-0301 to extract keywords/phrases from a target document, subsequently redacting $p\%$ of them. gpt-4-0613 then generates a query from this redacted document, guided solely by an example query format, without paired documents. (For ArguAna, we do not provide an example query format, since the LLM has sufficient prior knowledge of the argument and counterargument format.)

\begin{samepage}
\textbf{Prompt Instruction Regularization}: Refer to Figure \ref{fig:sigrid_pipline} (b) and Section \ref{ssec: prompt regularization}. We use gpt-4-0613 with a prompt designed to elicit the breakdown and paraphrasing of query aspects. All prompts are available in the project GitHub repository.\footnote{\url{https://github.com/Info-Regularization/Information-Regularization/tree/main/prompt}}
\end{samepage}

\textbf{Output Query Regularization}: gpt-3.5-turbo-0301 is used to summarize the original query into a simpler query. This method was applied to both document and instruction regularization outputs in our tests.

For all prompts used for keyword extraction and synthetic query generation, see Appendix \ref{sec: appendix_prompt_design}.

\subsection{Training Methods}
\label{ssec: training methods}

We fine-tuned each model on the different regularized datasets for one epoch, utilizing the AdamW optimizer \cite{adamw} with an initial learning rate of 1e-5, coupled with a cosine annealing schedule. While various warm-up ratios were tested, omitting the warm-up phase yielded more stable training/validation loss curves and enhanced performance. No additional hyperparameter tuning was done; hyperparameters were set to their default values. To maintain consistent training memory, texts exceeding 512 tokens were truncated to their initial 512 tokens, a sufficient range as most texts fell within this limit. We employed the standard NT-Xent loss with in-batch negatives, setting the temperature $\tau = 0.05$ and a batch size of 80, the maximum capacity for four 40G NVIDIA A100 GPUs. To minimize gradient noise during minibatch processing, we accumulated gradients across 20 minibatches.

The synthetically generated query $q_i$, the relevant document $d_i$, and the batch size $N$ are used as inputs to an embedding function $f$ with range in $\mathbb{R}^n$. The NT-Xent Loss \cite{simclr} is then defined as:
\begin{align}
\label{eq:ntxent}
L = \underset{{1 \leq i \leq N}} {\mathbb{E}} \left[-\log \frac{e^{\langle f(q_i), f(d_i) \rangle / \tau}}{\underset{1 \leq j \leq N}{\sum} e^{\langle f(q_i), f(d_j) \rangle / \tau}} \right]
\end{align}

All experiments were performed using 20 random seeds, with the averaged results shown in Tables \ref{tab: wtb table}, \ref{tab: doris-mae table} and \ref{tab: arguana table}. We use t-tests to compare the effectiveness of our regularization strategies against the 8-shot Promptagator baseline. Because multiple methods are being compared, we use Bonferroni correction \cite{bonferroni} with a correction
factor 8 (the number of IR2 methods). Using the adjusted p-value, colored arrows in the tables indicate statistically significant increases or decreases (adjusted $p< 0.05$) relative to the Promptagator baseline.

\begin{table*}[ht]
\setlength{\tabcolsep}{1.5pt}
\centering
\scriptsize
\begin{tabular}{p{2.5cm}p{1.55cm}p{1.55cm}p{1.55cm}p{1.55cm}|p{1.55cm}p{1.55cm}p{1.55cm}p{1.55cm}}

\toprule
& \multicolumn{4}{c}{E5-Large-v2} & \multicolumn{4}{c}{RoBERTa} \\
\cmidrule(lr){2-5} \cmidrule(lr){6-9}
Method & R@5 & R@20 & NDCG@10 & MRR@10  & R@5 & R@20 & NDCG@10 & MRR@10 \\
\midrule
Pretrained & 14.67 & 42.15 & 71.98 & 14.34 & 11.96 & 34.77 & 66.86 & 8.56\\
Contrastive Fine-tuning   & 15.28 & 39.87 & 72.51 & 13.43 & 14.75 & 43.32 & 73.55 & \textbf{20.79} \\
\cmidrule(lr){1-1}
Promptagator & 16.18 & 45.59 & 73.95 & 14.83 & 13.03 & 42.88 & 72.23 & 16.82 \\

\cmidrule(lr){1-1}

Doc\textsubscript{40\% reg} & 14.90 \color{red} $\downarrow$ -1.28  & 46.55 \color{darkgreen} $\uparrow$ 0.96  & 74.28 \color{darkgreen} $\uparrow$ 0.33  & 14.87 & 12.54 & 44.89 \color{darkgreen} $\uparrow$ 2.01  & 72.94 \color{darkgreen} $\uparrow$ 0.71  & 18.16 \color{darkgreen} $\uparrow$ 1.34  \\
Query\textsubscript{reg} $\circ$  Doc\textsubscript{40\% reg} & 15.85 & 45.67 & 73.96 & 16.10 \color{darkgreen} $\uparrow$ 1.27 & 14.66 \color{darkgreen} $\uparrow$ 1.63  & 47.41 \color{darkgreen} $\uparrow$ 4.53  & 74.04 \color{darkgreen} $\uparrow$ 1.81  & 18.81 \color{darkgreen} $\uparrow$ 1.99 \\

Doc\textsubscript{60\% reg}  & 15.23 \color{red} $\downarrow$ -0.95  & 46.57 \color{darkgreen} $\uparrow$ 0.98  & 74.98 \color{darkgreen} $\uparrow$ 1.03  & 15.14 & 14.11 \color{darkgreen} $\uparrow$ 1.08  & 43.99 \color{darkgreen} $\uparrow$ 1.11  & 72.94 \color{darkgreen} $\uparrow$ 0.71  & 16.84 \\

Query\textsubscript{reg} $\circ$  Doc\textsubscript{60\% reg} & 15.31 \color{red} $\downarrow$ -0.87  & 48.16 \color{darkgreen} $\uparrow$ 2.57  & 74.90 \color{darkgreen} $\uparrow$ 0.95  & 15.43 \color{darkgreen} $\uparrow$ 0.60 & 14.75 \color{darkgreen} $\uparrow$ 1.72  & 47.25 \color{darkgreen} $\uparrow$ 4.37  & 74.61 \color{darkgreen} $\uparrow$ 2.38  & 20.48 \color{darkgreen} $\uparrow$ 3.66    \\

Doc\textsubscript{80\% reg} &  14.68 \color{red} $\downarrow$ -1.50  & 43.89 \color{red} $\downarrow$ -1.70  & 74.13 \color{darkgreen} $\uparrow$ 0.18  & 13.51 \color{red} $\downarrow$ -1.32 & 14.47 \color{darkgreen} $\uparrow$ 1.44  & 42.47 & 72.61 \color{darkgreen} $\uparrow$ 0.38  & 15.10 \color{red} $\downarrow$ -1.72 \\

Query\textsubscript{reg} $\circ$  Doc\textsubscript{80\% reg} & 15.63 \color{red} $\downarrow$ -0.55  & 46.75 \color{darkgreen} $\uparrow$ 1.16  & 75.08 \color{darkgreen} $\uparrow$ 1.13  & 16.62 \color{darkgreen} $\uparrow$ 1.79 & \textbf{15.84} \color{darkgreen} $\uparrow$ 2.81  & 46.96 \color{darkgreen} $\uparrow$ 4.08  & \textbf{74.71} \color{darkgreen} $\uparrow$ 2.48  & 20.49 \color{darkgreen} $\uparrow$ 3.67 \\

\cmidrule(lr){1-1}

Instr\textsubscript{reg} & \textbf{16.38} & 46.73 \color{darkgreen} $\uparrow$ 1.14  & 74.42 \color{darkgreen} $\uparrow$ 0.47  & 17.06 \color{darkgreen} $\uparrow$ 2.23 & 13.80 \color{darkgreen} $\uparrow$ 0.77  & 45.78 \color{darkgreen} $\uparrow$ 2.90  & 73.53 \color{darkgreen} $\uparrow$ 1.30  & 18.24 \color{darkgreen} $\uparrow$ 1.42  \\

Query\textsubscript{reg} $\circ$  Instr\textsubscript{reg} & 15.76 \color{red} $\downarrow$ -0.42  & \textbf{48.46} \color{darkgreen} $\uparrow$ 2.87  & \textbf{76.02} \color{darkgreen} $\uparrow$ 2.07  & \textbf{20.86} \color{darkgreen} $\uparrow$ 6.03 & 15.20 \color{darkgreen} $\uparrow$ 2.17  & \textbf{47.44} \color{darkgreen} $\uparrow$ 4.56  & 74.08 \color{darkgreen} $\uparrow$ 1.85  & 19.31 \color{darkgreen} $\uparrow$ 2.49 \\
\midrule
& \multicolumn{4}{c}{SimCSE} & \multicolumn{4}{c}{SPECTER-v2} \\
\cmidrule(lr){2-5} \cmidrule(lr){6-9}
Method & R@5 & R@20 & NDCG@10 & MRR@10  & R@5 & R@20 & NDCG@10 & MRR@10 \\
\midrule
Pretrained & 14.38 & 41.83 & 70.81 & \textbf{24.83} & 13.59 & 41.92 & 71.46 & 21.74\\
Contrastive Fine-tuning & 15.60 & 44.99 & 73.18 & 22.41 & 14.56 & \textbf{45.64} & 71.46 & 20.49 \\
\cmidrule(lr){1-1}
Promptagator & 16.35 & 46.69 & 73.97 & 16.54 & 14.13 & 39.22 & 71.55 & 22.37\\
\cmidrule(lr){1-1}

Doc\textsubscript{40\% reg} & \textbf{17.70} \color{darkgreen} $\uparrow$ 1.35  & 45.30 \color{red} $\downarrow$ -1.39  & 74.49 \color{darkgreen} $\uparrow$ 0.52  & 16.49 & 15.39 \color{darkgreen} $\uparrow$ 1.26  & 43.39 \color{darkgreen} $\uparrow$ 4.17  & 72.11 \color{darkgreen} $\uparrow$ 0.56  & 21.66 \color{red} $\downarrow$ -0.71 \\

Query\textsubscript{reg} $\circ$ Doc\textsubscript{40\% reg} & 16.06 & 45.46 \color{red} $\downarrow$ -1.23  & 73.23 \color{red} $\downarrow$ -0.74  & 16.56 & 14.19 & 42.66 \color{darkgreen} $\uparrow$ 3.44  & 71.49 & 21.40 \color{red} $\downarrow$ -0.97  \\

Doc\textsubscript{60\% reg}  & 17.08 \color{darkgreen} $\uparrow$ 0.73  & 45.33 \color{red} $\downarrow$ -1.36  & 74.70 \color{darkgreen} $\uparrow$ 0.73  & 17.75 \color{darkgreen} $\uparrow$ 1.21 & 15.41 \color{darkgreen} $\uparrow$ 1.28  & 44.19 \color{darkgreen} $\uparrow$ 4.97  & 72.81 \color{darkgreen} $\uparrow$ 1.26  & 21.35 \color{red} $\downarrow$ -1.02  \\

Query\textsubscript{reg} $\circ$  Doc\textsubscript{60\% reg}  & 17.04 \color{darkgreen} $\uparrow$ 0.69  & 46.68 & 74.65 \color{darkgreen} $\uparrow$ 0.68  & 18.32 \color{darkgreen} $\uparrow$ 1.78 & 14.95 \color{darkgreen} $\uparrow$ 0.82  & 43.30 \color{darkgreen} $\uparrow$ 4.08  & 72.19 \color{darkgreen} $\uparrow$ 0.64  & 21.60 \color{red} $\downarrow$ -0.77  \\

Doc\textsubscript{80\% reg} & 15.01 \color{red} $\downarrow$ -1.34  & 42.53 \color{red} $\downarrow$ -4.16  & 73.69 \color{red} $\downarrow$ -0.28  & 18.68 \color{darkgreen} $\uparrow$ 2.14 & \textbf{17.02} \color{darkgreen} $\uparrow$ 2.89  & 44.97 \color{darkgreen} $\uparrow$ 5.75  & \textbf{73.94} \color{darkgreen} $\uparrow$ 2.39  & \textbf{22.78} \color{darkgreen} $\uparrow$ 0.41   \\

Query\textsubscript{reg} $\circ$  Doc\textsubscript{80\% reg} & 15.92 & 45.70 \color{red} $\downarrow$ -0.99  & 73.86 & 18.54 \color{darkgreen} $\uparrow$ 2.00 & 14.70 \color{darkgreen} $\uparrow$ 0.57  & 44.66 \color{darkgreen} $\uparrow$ 5.44  & 73.07 \color{darkgreen} $\uparrow$ 1.52  & 22.65  \\

\cmidrule(lr){1-1}

Instr\textsubscript{reg} & 16.50 & 46.68 & \textbf{75.60} \color{darkgreen} $\uparrow$ 1.63  & 19.71 \color{darkgreen} $\uparrow$ 3.17 & 14.73 \color{darkgreen} $\uparrow$ 0.60  & 41.50 \color{darkgreen} $\uparrow$ 2.28  & 73.06 \color{darkgreen} $\uparrow$ 1.51  & 21.05 \color{red} $\downarrow$ -1.32   \\

Query\textsubscript{reg} $\circ$ Instr\textsubscript{reg} & 16.06 & \textbf{47.61} \color{darkgreen} $\uparrow$ 0.92  & 75.43 \color{darkgreen} $\uparrow$ 1.46  & 22.14 \color{darkgreen} $\uparrow$ 5.60 & 14.50 \color{darkgreen} $\uparrow$ 0.37  & 41.44 \color{darkgreen} $\uparrow$ 2.22  & 72.46 \color{darkgreen} $\uparrow$ 0.91  & 22.26 \\

\bottomrule
\end{tabular}
\caption{Result for DORIS-MAE. See Table \ref{tab: wtb table} caption for reporting conventions.}
\label{tab: doris-mae table}
\end{table*}

\section{Results}

\label{sec: results}

\subsection{Results on WhatsThatBook}
\label{ssec: results on whatsthatbook}

Table \ref{tab: wtb table} shows comparisons between the IR2 methods and the Promptagator baseline. Except for Doc\textsubscript{80\% reg}, RoBERTa and SPECTER-v2 show consistent improvement with all regularization methods. For E5 and SimCSE, there is no clear signal of improvement from document or instruction regularization alone.
However, query regularization, combined with either document or instruction regularization, improves performance across all models. 

Not all models respond to fine-tuning on synthetic datasets in the same way. SPECTER-v2 and RoBERTa start from similar performance levels, but RoBERTa reaches considerably higher performance after fine-tuning. This suggests that the effectiveness of synthetic data augmentation is model-dependent. The SPECTER-v2 checkpoint had been previously fine-tuned for scientific document understanding, possibly interfering with its performance on the WhatsThatBook dataset. 

Document regularization with an 80\% masking ratio leads to synthetic queries that are either too vague or contain misleading and false information. Consequently, for E5, RoBERTa, and SimCSE, we observed that document regularization with an 80\% mask resulted in significantly lower performances compared to Promptagator. Results are stronger with a masking level between 40\% and 60\%.

\subsection{Results on DORIS-MAE}
\label{ssec: results on doris-mae}

Table \ref{tab: doris-mae table} shows that instruction regularization, alone or combined with query regularization, consistently enhances performance across various models and metrics, surpassing the Promptagator's synthetic dataset, with minor exceptions within a 1\% margin. The strongest results across all models and data synthesis methods are achieved by query regularization combined with instruction regularization.

Document regularization has variable performance depending on the mask percentage ($p=40\%, 60\%, 80\%$) and model. It has only a small effect on the performance of E5-Large-v2 compared to the Promptagator baseline, and underperforms with SimCSE. However, it substantially improves performance when applied to RoBERTa and SPECTER-v2. 

For E5, SimCSE, and RoBERTa, integrating query regularization with document or instruction regularization improves performance across all metrics. Query regularization decreases the textual similarity between queries and their target documents, suggesting that models are learning deeper semantic relationships between queries and documents.

\subsection{Results on ArguAna}
\label{ssec: results on arguana}

\begin{table*}[!ht]
\centering
\setlength{\tabcolsep}{1.5pt}
\scriptsize
\begin{tabular}{p{2.5cm}p{1.55cm}p{1.55cm}p{1.55cm}p{1.55cm}|p{1.55cm}p{1.55cm}p{1.55cm}p{1.55cm}}
\toprule
& \multicolumn{4}{c}{E5-Large-v2} & \multicolumn{4}{c}{RoBERTa} \\
\cmidrule(lr){2-5} \cmidrule(lr){6-9}
Method & R@5 & R@20 & NDCG@10 & MRR@10  & R@5 & R@20 & NDCG@10 & MRR@10 \\
\midrule
Pretrained & 59.37 & 88.16 & 47.75 & 38.96 & 30.51 & 49.68 & 23.56 & 18.49 \\
Contrastive Fine-tuning   & 70.69 & 94.77 & 56.15 & 47.13 & 59.12 & 84.67 & 47.57 & 39.65 \\
\cmidrule(lr){1-1}
Promptagator & 69.41 & 94.11 & 55.14 & 46.67 & 61.51 & 88.19 & 49.80 & 41.85 \\

\cmidrule(lr){1-1}

Doc\textsubscript{40\% reg} & \textbf{78.72} \color{darkgreen} $\uparrow$ 9.31  & \textbf{96.46} \color{darkgreen} $\uparrow$ 2.35  & 64.24 \color{darkgreen} $\uparrow$ 9.10  & 56.45 \color{darkgreen} $\uparrow$ 9.78 & 67.38 \color{darkgreen} $\uparrow$ 5.87  & 91.03 \color{darkgreen} $\uparrow$ 2.84  & 54.54 \color{darkgreen} $\uparrow$ 4.74  & 46.43 \color{darkgreen} $\uparrow$ 4.58  \\
Query\textsubscript{reg} $\circ$  Doc\textsubscript{40\% reg} & 77.76 \color{darkgreen} $\uparrow$ 8.35  & 96.21 \color{darkgreen} $\uparrow$ 2.10  & 63.00 \color{darkgreen} $\uparrow$ 7.86  & 55.24 \color{darkgreen} $\uparrow$ 8.57 & 67.29 \color{darkgreen} $\uparrow$ 5.78  & \textbf{91.08} \color{darkgreen} $\uparrow$ 2.89  & 54.04 \color{darkgreen} $\uparrow$ 4.24  & 45.85 \color{darkgreen} $\uparrow$ 4.00 \\

Doc\textsubscript{60\% reg}  & 78.16 \color{darkgreen} $\uparrow$ 8.75  & 96.41 \color{darkgreen} $\uparrow$ 2.30  & \textbf{64.41} \color{darkgreen} $\uparrow$ 9.27  & \textbf{56.57} \color{darkgreen} $\uparrow$ 9.90 & 67.51 \color{darkgreen} $\uparrow$ 6.00  & 90.77 \color{darkgreen} $\uparrow$ 2.58  & \textbf{54.85} \color{darkgreen} $\uparrow$ 5.05  & \textbf{46.85} \color{darkgreen} $\uparrow$ 5.00 \\

Query\textsubscript{reg} $\circ$  Doc\textsubscript{60\% reg} &  78.34 \color{darkgreen} $\uparrow$ 8.93  & 96.37 \color{darkgreen} $\uparrow$ 2.26  & 64.10 \color{darkgreen} $\uparrow$ 8.96  & 56.24 \color{darkgreen} $\uparrow$ 9.57 & 67.71 \color{darkgreen} $\uparrow$ 6.20  & 90.99 \color{darkgreen} $\uparrow$ 2.80  & 54.41 \color{darkgreen} $\uparrow$ 4.61  & 46.34 \color{darkgreen} $\uparrow$ 4.49 \\

Doc\textsubscript{80\% reg} & 77.37 \color{darkgreen} $\uparrow$ 7.96  & 95.57 \color{darkgreen} $\uparrow$ 1.46  & 63.14 \color{darkgreen} $\uparrow$ 8.00  & 55.16 \color{darkgreen} $\uparrow$ 8.49 & \textbf{67.82} \color{darkgreen} $\uparrow$ 6.31  & 90.18 \color{darkgreen} $\uparrow$ 1.99  & 54.62 \color{darkgreen} $\uparrow$ 4.82  & 46.79 \color{darkgreen} $\uparrow$ 4.94  \\

Query\textsubscript{reg} $\circ$  Doc\textsubscript{80\% reg} & 78.44 \color{darkgreen} $\uparrow$ 9.03  & 96.25 \color{darkgreen} $\uparrow$ 2.14  & 64.36 \color{darkgreen} $\uparrow$ 9.22  & 56.49 \color{darkgreen} $\uparrow$ 9.82 & 67.38 \color{darkgreen} $\uparrow$ 5.87  & 90.70 \color{darkgreen} $\uparrow$ 2.51  & 54.53 \color{darkgreen} $\uparrow$ 4.73  & 46.59 \color{darkgreen} $\uparrow$ 4.74  \\

\cmidrule(lr){1-1}

Instr\textsubscript{reg} & 72.06 \color{darkgreen} $\uparrow$ 2.65  & 95.70 \color{darkgreen} $\uparrow$ 1.59  & 57.76 \color{darkgreen} $\uparrow$ 2.62  & 49.17 \color{darkgreen} $\uparrow$ 2.50 & 64.47 \color{darkgreen} $\uparrow$ 2.96  & 90.09 \color{darkgreen} $\uparrow$ 1.90  & 51.49 \color{darkgreen} $\uparrow$ 1.69  & 43.20 \color{darkgreen} $\uparrow$ 1.35  \\

Query\textsubscript{reg} $\circ$  Instr\textsubscript{reg} &  71.59 \color{darkgreen} $\uparrow$ 2.18  & 95.18 \color{darkgreen} $\uparrow$ 1.07  & 56.89 \color{darkgreen} $\uparrow$ 1.75  & 48.37 \color{darkgreen} $\uparrow$ 1.70 & 65.14 \color{darkgreen} $\uparrow$ 3.63  & 90.26 \color{darkgreen} $\uparrow$ 2.07  & 52.25 \color{darkgreen} $\uparrow$ 2.45  & 43.89 \color{darkgreen} $\uparrow$ 2.04 \\
\midrule
& \multicolumn{4}{c}{SimCSE} & \multicolumn{4}{c}{SPECTER-v2} \\
\cmidrule(lr){2-5} \cmidrule(lr){6-9}
Method & R@5 & R@20 & NDCG@10 & MRR@10  & R@5 & R@20 & NDCG@10 & MRR@10 \\
\midrule
Pretrained & 49.03 & 81.12 & 39.23 & 30.35 & 38.91 & 71.79 & 30.91 & 23.23\\
Contrastive Fine-tuning & 36.70 & 64.79 & 28.85 & 22.12 & 41.90 & 75.70 & 33.20 & 25.26\\
\cmidrule(lr){1-1}
Promptagator & 69.49 & 92.54 & 56.38 & 47.97 & 38.64 & 75.15 & 30.24 & 21.77 \\
\cmidrule(lr){1-1}

Doc\textsubscript{40\% reg} & 74.92 \color{darkgreen} $\uparrow$ 5.43  & 93.36 \color{darkgreen} $\uparrow$ 0.82  & 60.24 \color{darkgreen} $\uparrow$ 3.86  & 52.42 \color{darkgreen} $\uparrow$ 4.45  & 56.09 \color{darkgreen} $\uparrow$ 17.45  & 86.32 \color{darkgreen} $\uparrow$ 11.17  & 45.00 \color{darkgreen} $\uparrow$ 14.76  & 36.30 \color{darkgreen} $\uparrow$ 14.53\\

Query\textsubscript{reg} $\circ$ Doc\textsubscript{40\% reg} &  \textbf{76.07} \color{darkgreen} $\uparrow$ 6.58  & \textbf{93.97} \color{darkgreen} $\uparrow$ 1.43  & \textbf{61.38} \color{darkgreen} $\uparrow$ 5.00  & \textbf{53.57} \color{darkgreen} $\uparrow$ 5.60 & 51.28 \color{darkgreen} $\uparrow$ 12.64  & 83.33 \color{darkgreen} $\uparrow$ 8.18  & 40.59 \color{darkgreen} $\uparrow$ 10.35  & 31.81 \color{darkgreen} $\uparrow$ 10.04 \\

Doc\textsubscript{60\% reg}  &  73.20 \color{darkgreen} $\uparrow$ 3.71  & 92.10 & 58.46 \color{darkgreen} $\uparrow$ 2.08  & 50.58 \color{darkgreen} $\uparrow$ 2.61 & 58.87 \color{darkgreen} $\uparrow$ 20.23  & 86.93 \color{darkgreen} $\uparrow$ 11.78  & 46.55 \color{darkgreen} $\uparrow$ 16.31  & 37.88 \color{darkgreen} $\uparrow$ 16.11 \\

Query\textsubscript{reg} $\circ$  Doc\textsubscript{60\% reg}  & 75.75 \color{darkgreen} $\uparrow$ 6.26  & 93.19 \color{darkgreen} $\uparrow$ 0.65  & 60.94 \color{darkgreen} $\uparrow$ 4.56  & 53.24 \color{darkgreen} $\uparrow$ 5.27 & 54.22 \color{darkgreen} $\uparrow$ 15.58  & 84.97 \color{darkgreen} $\uparrow$ 9.82  & 42.90 \color{darkgreen} $\uparrow$ 12.66  & 34.21 \color{darkgreen} $\uparrow$ 12.44 \\

Doc\textsubscript{80\% reg} &  70.38 & 89.67 \color{red} $\downarrow$ -2.87  & 55.76 & 47.97 & \textbf{61.02} \color{darkgreen} $\uparrow$ 22.38  & \textbf{87.48} \color{darkgreen} $\uparrow$ 12.33  & \textbf{48.68} \color{darkgreen} $\uparrow$ 18.44  & \textbf{40.24} \color{darkgreen} $\uparrow$ 18.47 \\

Query\textsubscript{reg} $\circ$  Doc\textsubscript{80\% reg} & 74.10 \color{darkgreen} $\uparrow$ 4.61  & 92.59 & 59.79 \color{darkgreen} $\uparrow$ 3.41  & 52.18 \color{darkgreen} $\uparrow$ 4.21 & 57.55 \color{darkgreen} $\uparrow$ 18.91  & 86.65 \color{darkgreen} $\uparrow$ 11.50  & 46.01 \color{darkgreen} $\uparrow$ 15.77  & 37.44 \color{darkgreen} $\uparrow$ 15.67 \\

\cmidrule(lr){1-1}

Instr\textsubscript{reg} & 70.97 & 93.91 \color{darkgreen} $\uparrow$ 1.37  & 57.41 & 48.80 & 43.26 \color{darkgreen} $\uparrow$ 4.62  & 80.16 \color{darkgreen} $\uparrow$ 5.01  & 34.80 \color{darkgreen} $\uparrow$ 4.56  & 26.07 \color{darkgreen} $\uparrow$ 4.30  \\

Query\textsubscript{reg} $\circ$ Instr\textsubscript{reg} & 71.46 \color{darkgreen} $\uparrow$ 1.97  & 93.78 \color{darkgreen} $\uparrow$ 1.24  & 58.16 \color{darkgreen} $\uparrow$ 1.78  & 49.78 \color{darkgreen} $\uparrow$ 1.81 & 38.57 & 77.12 \color{darkgreen} $\uparrow$ 1.97  & 31.72 \color{darkgreen} $\uparrow$ 1.48  & 23.15 \color{darkgreen} $\uparrow$ 1.38 \\

\bottomrule
\end{tabular}
\caption{Results for ArguAna. See Table \ref{tab: wtb table} caption for reporting conventions.}
\label{tab: arguana table}
\end{table*}

In Table \ref{tab: arguana table}, we observe that the regularized synthetic datasets consistently outperform baseline approaches. While instruction regularization continues to outperform baselines, all models achieve the strongest performance when fine-tuning on document regularized dataset. Similar to our observations in DORIS-MAE and WTB, query regularization, when combined with another synthetic query method, remains consistently strong.

\subsection{Analysis and Discussion}
\label{ssec: analysis and discussion}

Our experiments demonstrate substantial improvements from regularizing synthetic queries in IR tasks. Rather than simply altering task difficulty, these methods refine the model's ability to discern and process relevant information, an advancement over existing strategies. Specifically, integrating query and instruction regularization emerges as a robust method, enhancing the model's performance consistently.

Models trained with the proposed regularized synthetic data generation strategies not only outperform pretrained counterparts but also exhibit considerable gains over existing synthetic data methods like Promptagator. This is shown across various metrics and models.

The practical implications of these methods extend beyond improved model performance. They point toward a greater adaptability in handling complex information retrieval tasks, a critical component given the variability of real-world applications. Moreover, the insights gained from this research contribute to the broader discourse on the role of synthetic data in model training, especially in scenarios where data scarcity is a challenge.

Though IR2 improves over prior methods, there are several caveats. The document regularization method, which masks part of the input document and generates a query from this masked document, sometimes results in hallucinations in the queries. We hypothesize that performance differences across datasets are due to varying task sensitivity to these hallucinations. Performance was generally weaker with a mask ratio above 60\%. 

As shown in Table \ref{tab: wtb table}, document regularization had weaker performance on the WhatsThatBook dataset. This task requires a match between fine-grained details in the query and document. In contrast, for the ArguAna dataset in Table \ref{tab: arguana table}, where the task involves pairing arguments with counterarguments, the method had stronger performance. Thematic similarity is sufficient for this task.

\subsection{Cost Analysis}

Instruction Regularization, which costs \$100-200 per dataset, is the most expensive of the three regularization methods. Due to longer prompts with the inclusion of 8 example pairs, the baseline method Promptagator costs \$400-\$600 per dataset.

We attempted to generate synthetic queries by hand to estimate human labor costs, and required at least 5 minutes per query. Assuming a wage of \$15 per hour, human generation costs at least \$5,000 per dataset. This suggests substantial savings by using GPT-4.

\section{Limitations}

The effectiveness of our regularization strategies does show variability across different models, signaling that these methods may be more suited to certain architectures or training setups. Future work could delve deeper into this aspect, aiming to identify the underpinning factors that contribute to these discrepancies.



\twocolumn[\clearpage]

\bibliography{custom}

\newpage
\twocolumn[\clearpage]

\appendix

\section{Prompt Design}
\label{sec: appendix_prompt_design}
\subsection{DORIS-MAE Prompt}
\begin{figure*}[ht]
    \centering
        \includegraphics[width=\linewidth]{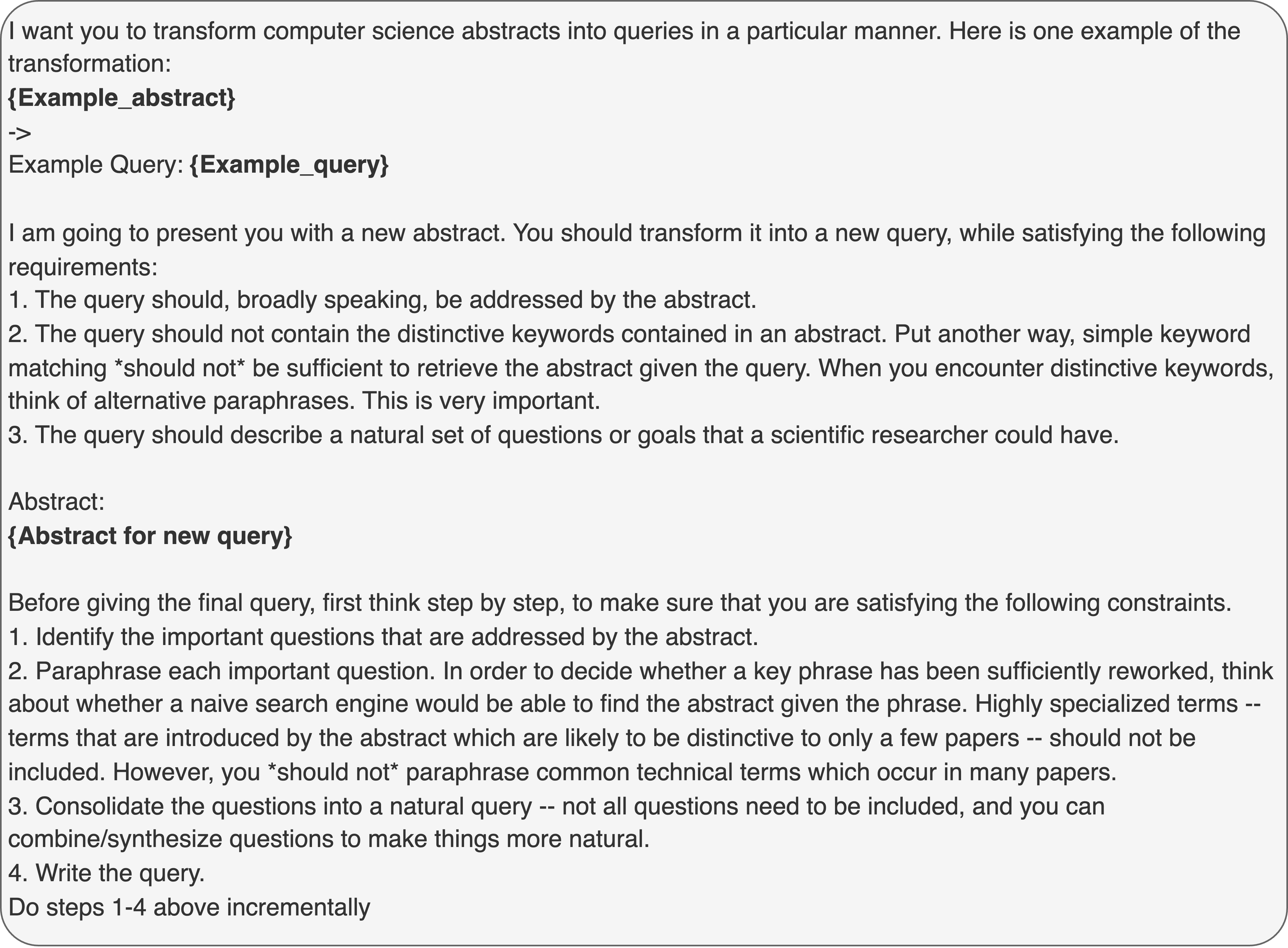}
    \caption{DORIS-MAE Instruction Regularization Prompt}
    \label{fig: dm_I_reg_prompt}
\end{figure*}

\begin{figure*}[ht]
    \centering
        \includegraphics[width=\linewidth]{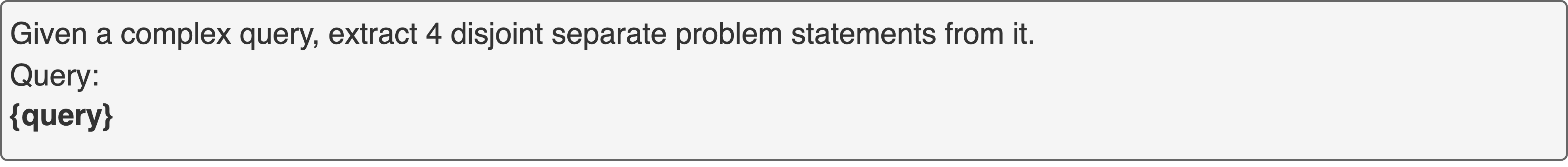}
    \caption{DORIS-MAE Query Regularization on Instruction Regularization Prompt}
    \label{fig: dm_q_reg_I_reg_prompt}
\end{figure*}

\begin{figure*}[ht]
    \centering
        \includegraphics[width=\linewidth]{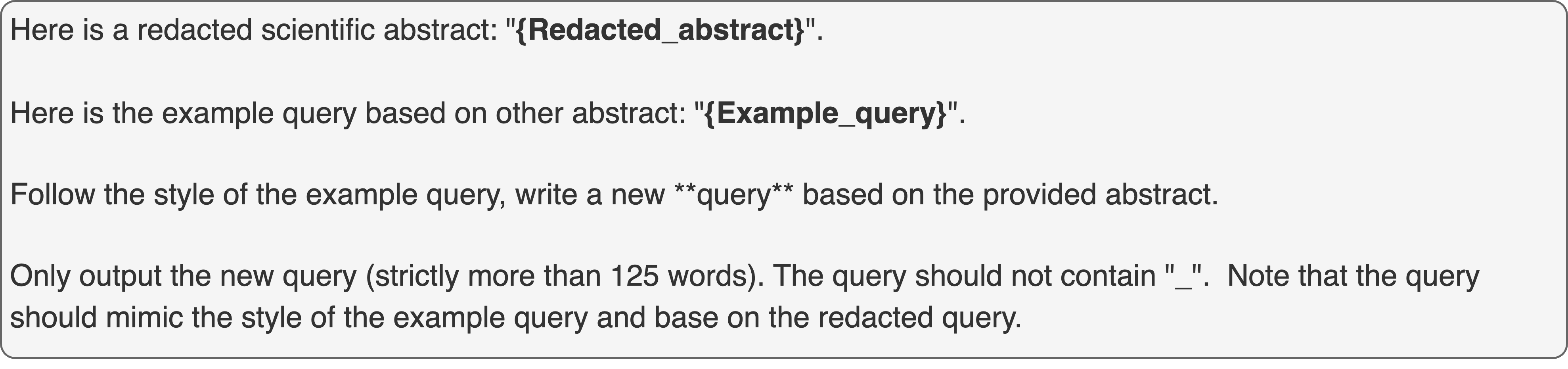}
    \caption{DORIS-MAE Document Regularization 40\%, 60\% Prompt}
    \label{fig: dm_d_reg_40_60_prompt}
\end{figure*}

\begin{figure*}[ht]
    \centering
        \includegraphics[width=\linewidth]{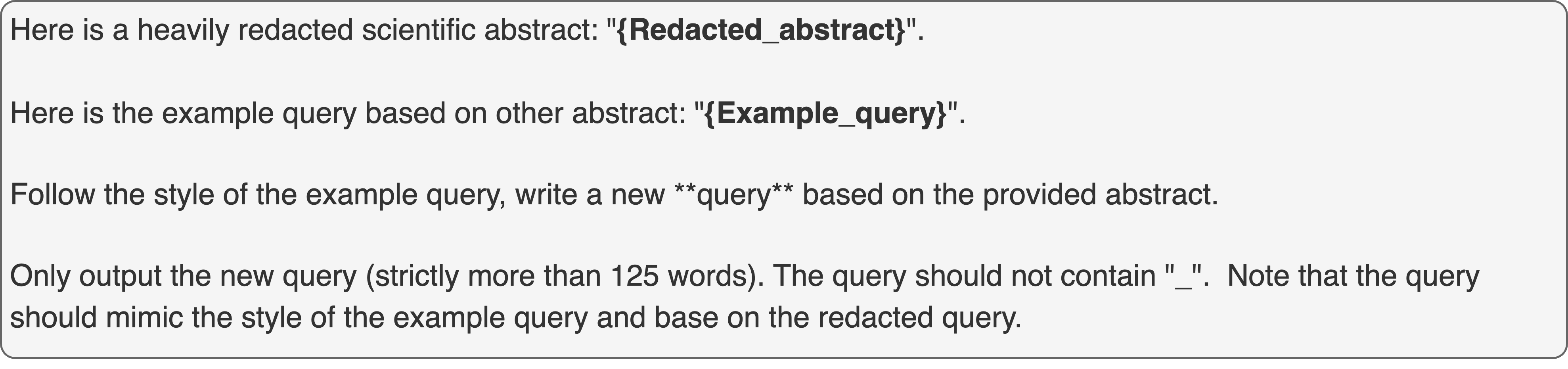}
    \caption{DORIS-MAE Document Regularization 80\% Prompt}
    \label{fig: dm_d_reg_80_prompt}
\end{figure*}

\begin{figure*}[ht]
    \centering
        \includegraphics[width=\linewidth]{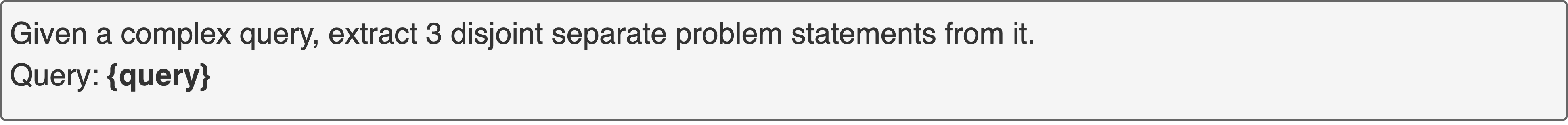}
    \caption{DORIS-MAE Query Regularization on synthetic queries from Doc Reg 40\%, 60\%, and 80\%}
    \label{fig: dm_q_reg_d_reg_prompt}
\end{figure*}

\begin{figure*}[ht]
    \centering
        \includegraphics[width=\linewidth]{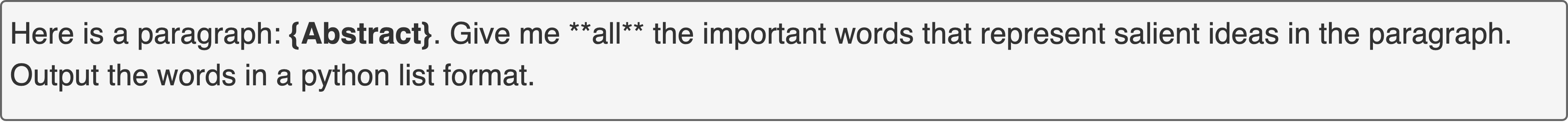}
    \caption{DORIS-MAE Keywords Extraction Prompt}
    \label{fig: dm_keyword_prompt}
\end{figure*}

\begin{figure*}[ht]
    \centering
        \includegraphics[width=\linewidth]{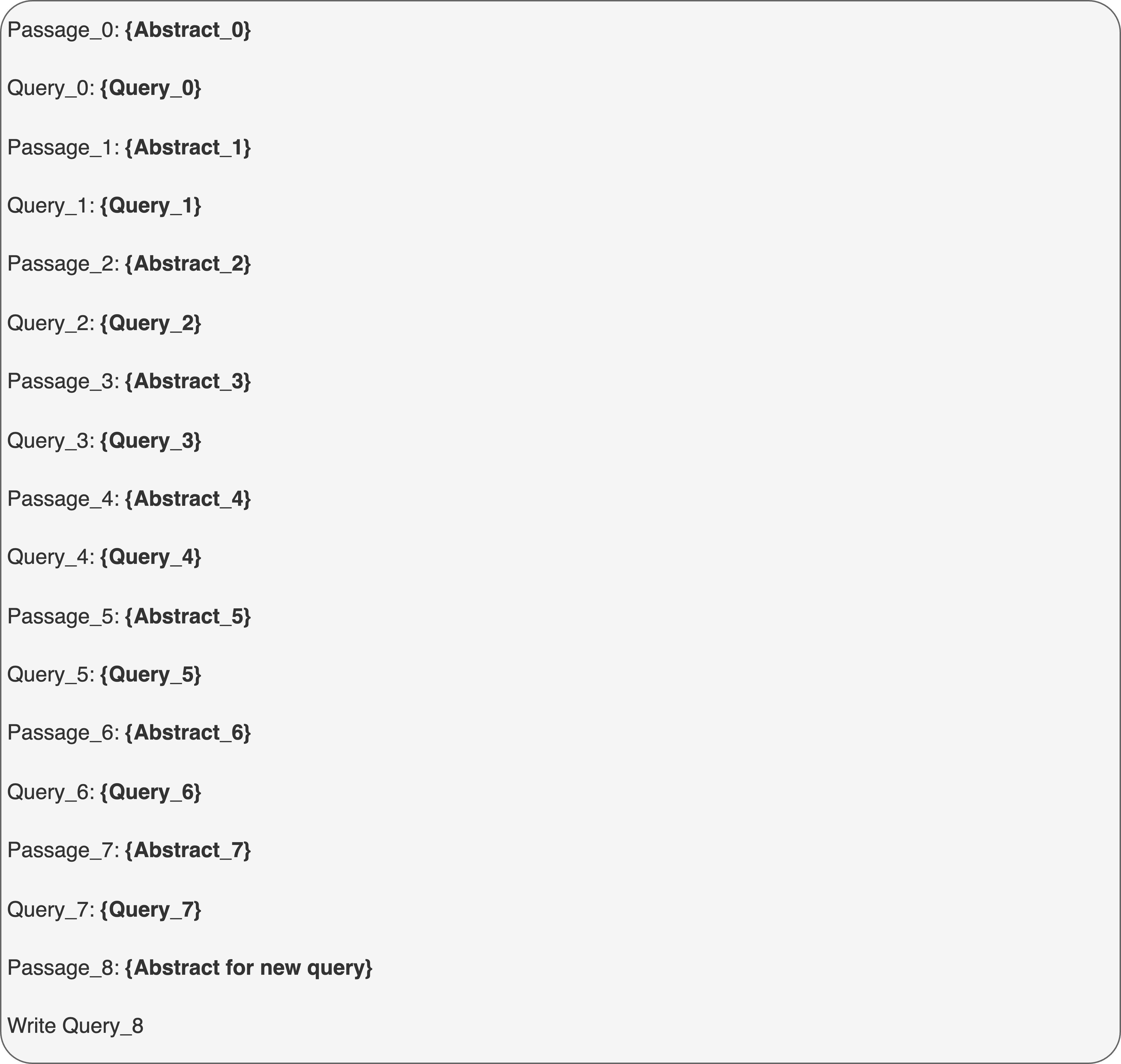}
    \caption{DORIS-MAE Promptagator Prompt}
    \label{fig: dm_promptagator_prompt}
\end{figure*}

Note, Query Regularization for DORIS-MAE will generate 4 shorter synthetic queries per document, we randomly choose one for later training. 
\begin{itemize}
    \item Figure \ref{fig: dm_I_reg_prompt} for Instruction Regularization prompt
    \item Figure \ref{fig: dm_q_reg_I_reg_prompt} for Query Regularization on Instruction Regularization prompt
    \item Figure \ref{fig: dm_d_reg_40_60_prompt} for Document Regularization 40\%, 60\% prompt
    \item Figure \ref{fig: dm_d_reg_80_prompt} for Document Regularization 80\% prompt
    \item Figure \ref{fig: dm_q_reg_d_reg_prompt} for Query Regularization on synthetic queries from Doc Reg 40\%, 60\%, and 80\%
    \item Figure \ref{fig: dm_keyword_prompt} for keywords extraction prompt
    \item Figure \ref{fig: dm_promptagator_prompt} for Promptagator prompt example
\end{itemize}

\subsection{ArguAna Prompt}
\begin{figure*}[ht]
    \centering
        \includegraphics[width=\linewidth]{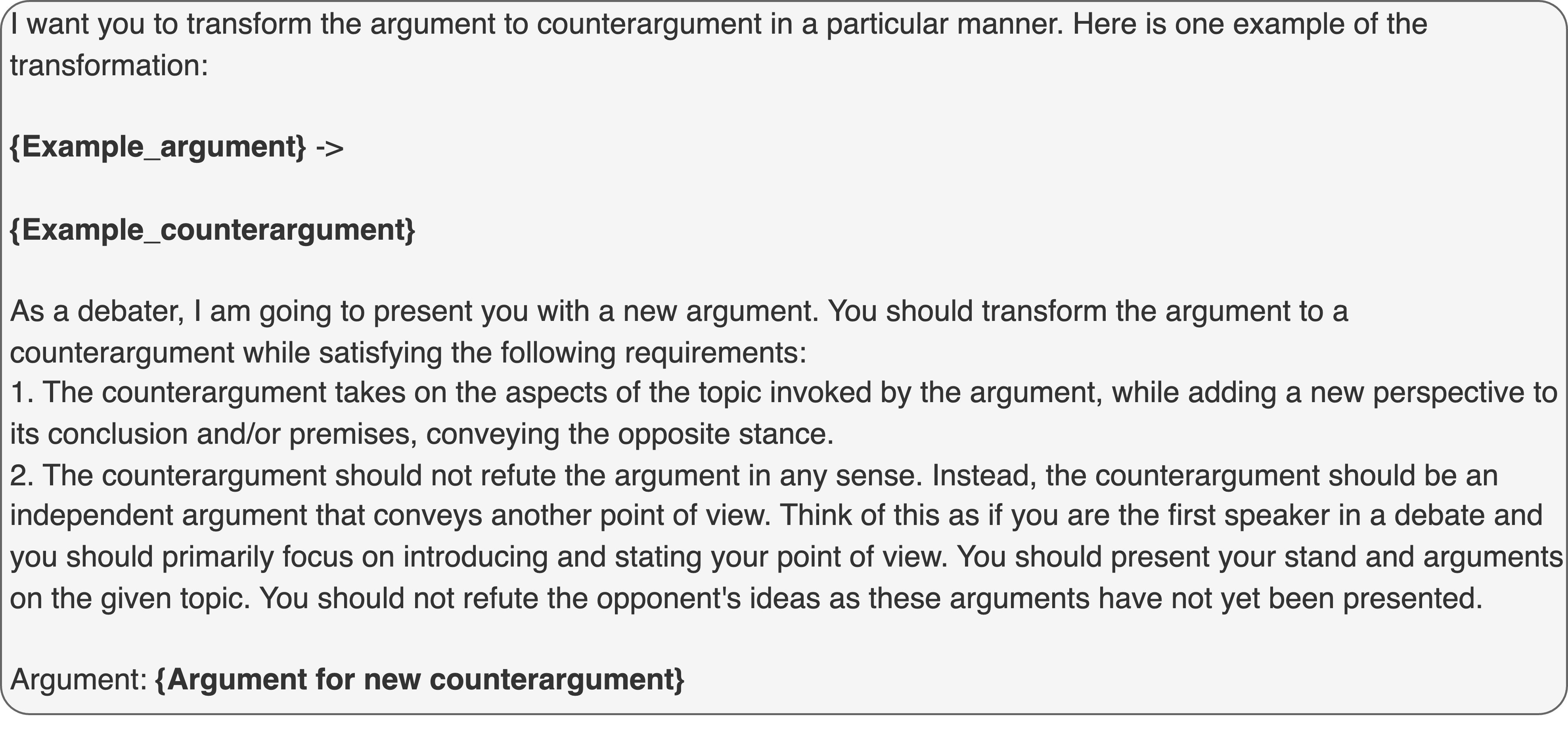}
    \caption{ArguAna Instruction Regularization Prompt}
    \label{fig: arguana_I_reg_prompt}
\end{figure*}

\begin{figure*}[ht]
    \centering
        \includegraphics[width=\linewidth]{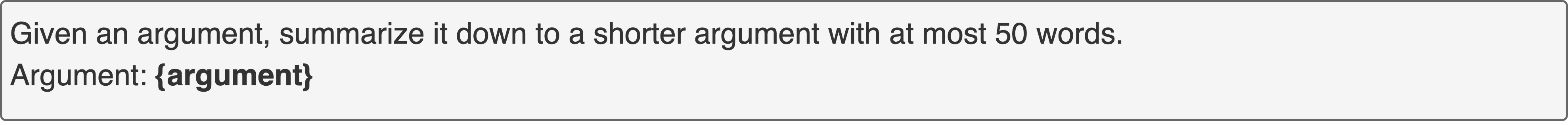}
    \caption{ArguAna Query Regularization on Instruction Regularization Prompt}
    \label{fig: arguana_q_reg_I_reg_prompt}
\end{figure*}

\begin{figure*}[ht]
    \centering
        \includegraphics[width=\linewidth]{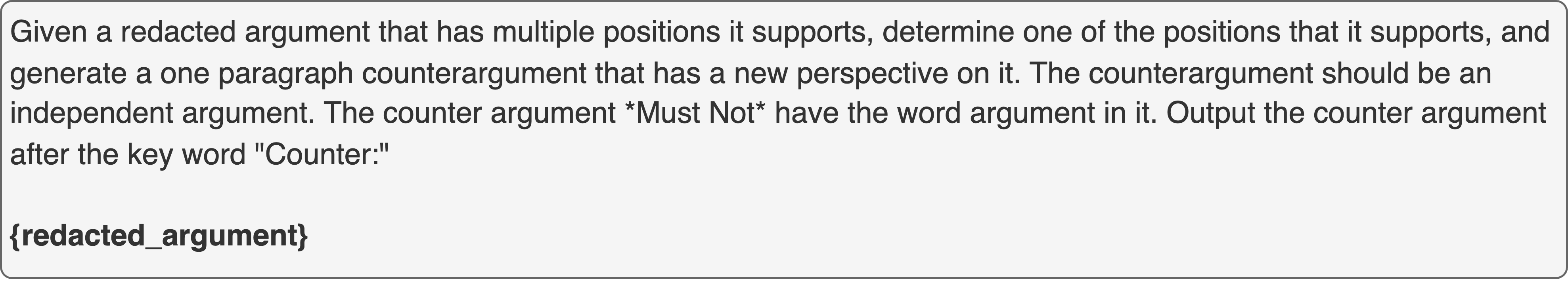}
    \caption{ArguAna Document Regularization 40\%, 60\%, and 80\% Prompt}
    \label{fig: arguana_d_reg_prompt}
\end{figure*}

\begin{figure*}[ht]
    \centering
        \includegraphics[width=\linewidth]{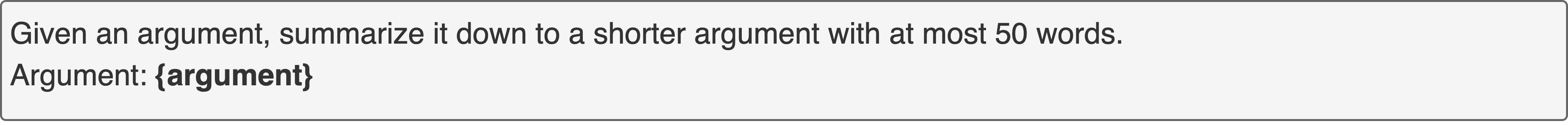}
    \caption{ArguAna Query Regularization on synthetic queries from Doc Reg 40\%, 60\%, and 80\%}
    \label{fig: arguana_q_reg_d_reg_prompt}
\end{figure*}

\begin{figure*}[ht]
    \centering
        \includegraphics[width=\linewidth]{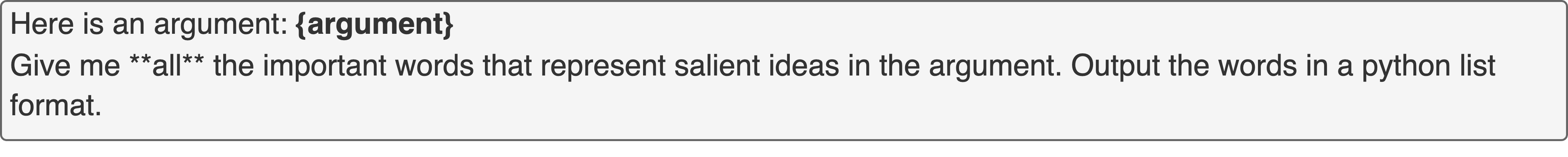}
    \caption{ArguAna Keywords Extraction Prompt}
    \label{fig: arguana_keyword_prompt}
\end{figure*}

\begin{figure*}[ht]
    \centering
        \includegraphics[width=\linewidth]{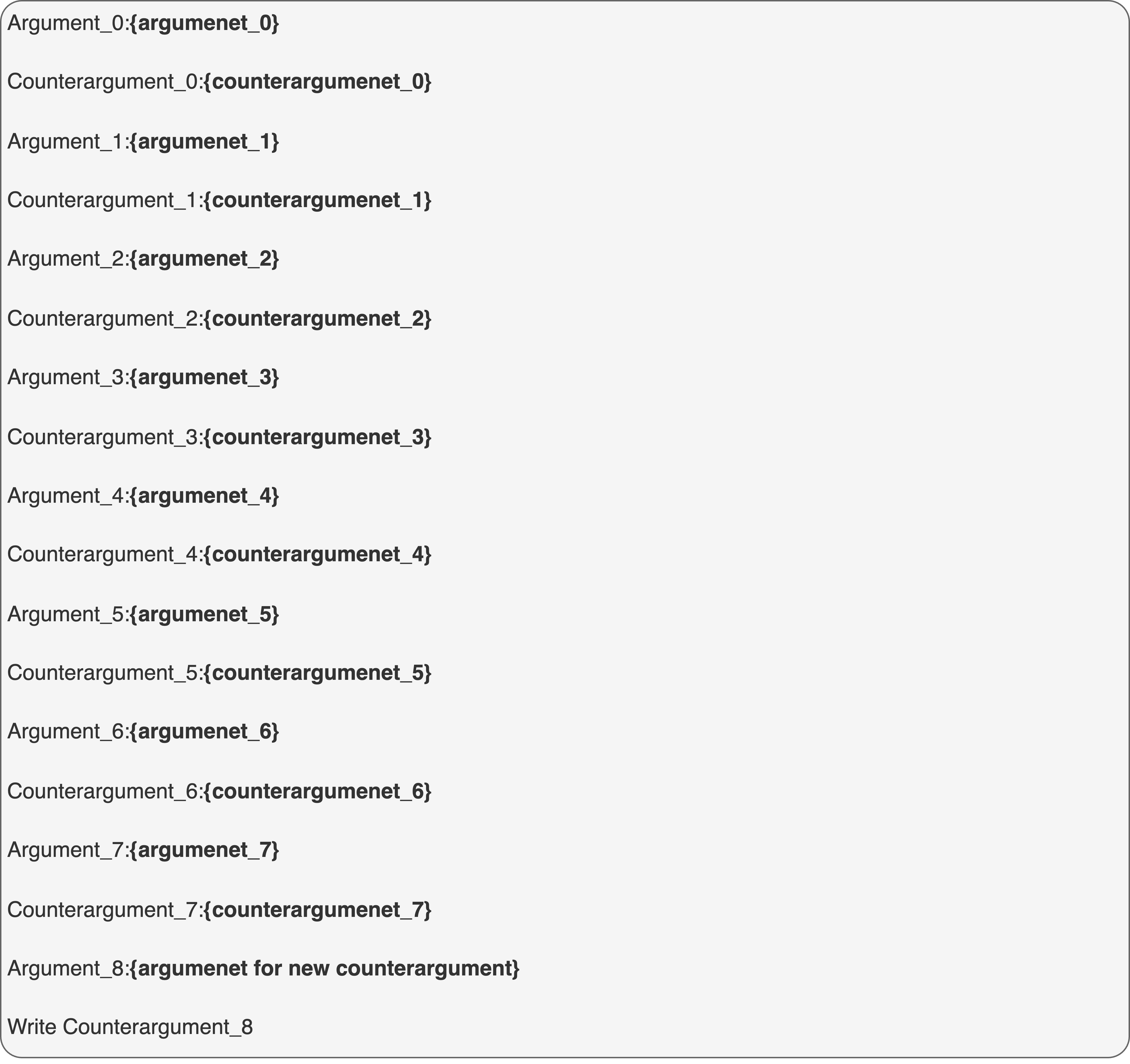}
    \caption{ArguAna Promptagator Prompt}
    \label{fig: arguana_promptagator_prompt}
\end{figure*}

\begin{itemize}
    \item Figure \ref{fig: arguana_I_reg_prompt} for Instruction Regularization prompt
    \item Figure \ref{fig: arguana_q_reg_I_reg_prompt} for Query Regularization on Instruction Regularization prompt
    \item Figure \ref{fig: arguana_d_reg_prompt} for Document Regularization 40\%, 60\%, and 80\% prompt
    \item Figure \ref{fig: arguana_q_reg_d_reg_prompt} for Query Regularization on synthetic queries from Doc Reg 40\%, 60\%, and 80\%
    \item Figure \ref{fig: arguana_keyword_prompt} for keywords extraction prompt
    \item Figure \ref{fig: arguana_promptagator_prompt} for Promptagator prompt example
\end{itemize}

\subsection{WhatsThatBook Prompt}
\begin{figure*}[ht]
    \centering
        \includegraphics[width=\linewidth]{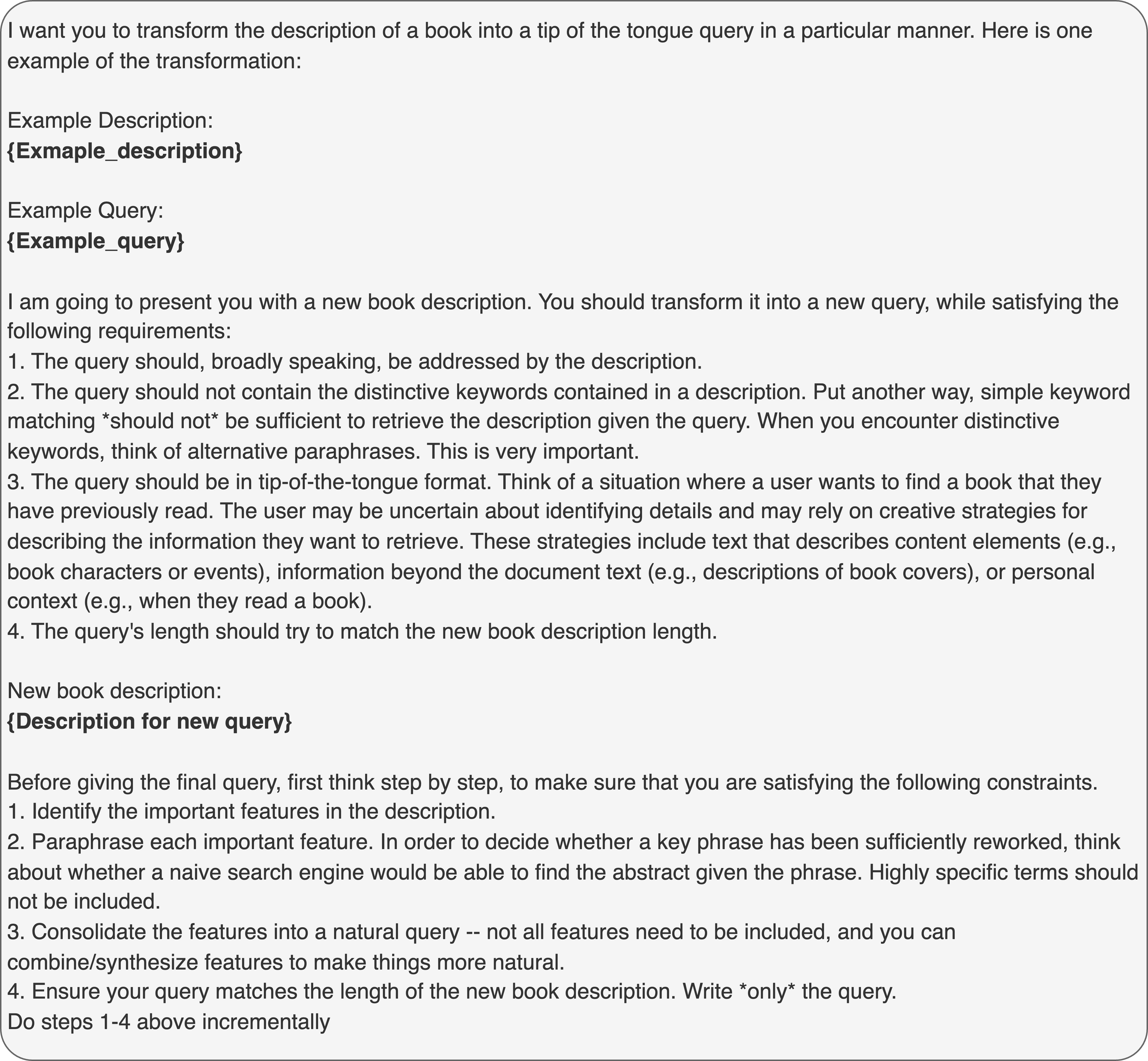}
    \caption{WhatsThatBook Instruction Regularization Prompt}
    \label{fig: wtb_I_reg_prompt}
\end{figure*}

\begin{figure*}[ht]
    \centering
        \includegraphics[width=\linewidth]{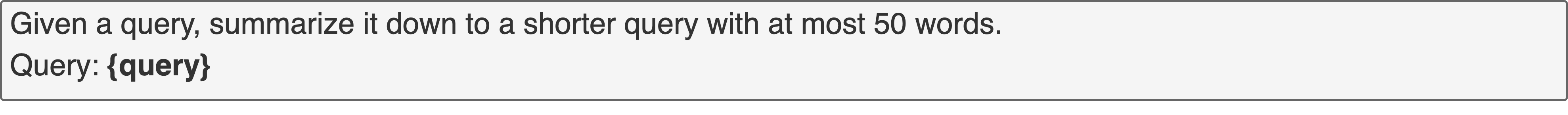}
    \caption{WhatsThatBook Query Regularization on Instruction Regularization Prompt}
    \label{fig: wtb_q_reg_I_reg_prompt}
\end{figure*}

\begin{figure*}[ht]
    \centering
        \includegraphics[width=\linewidth]{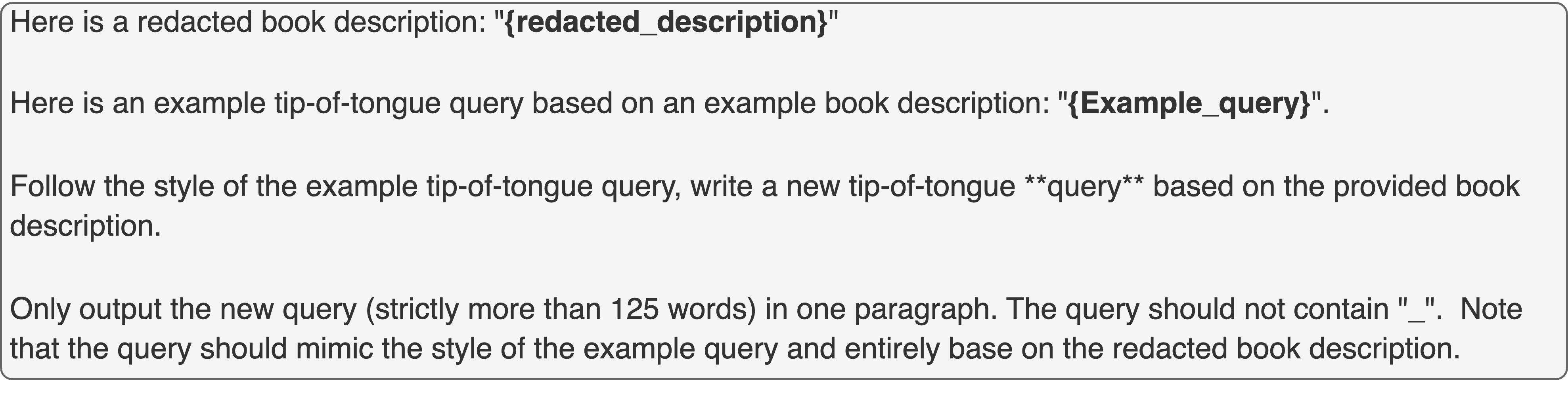}
    \caption{WhatsThatBook Document Regularization 40\% Prompt}
    \label{fig: wtb_d_reg_40_prompt}
\end{figure*}

\begin{figure*}[ht]
    \centering
        \includegraphics[width=\linewidth]{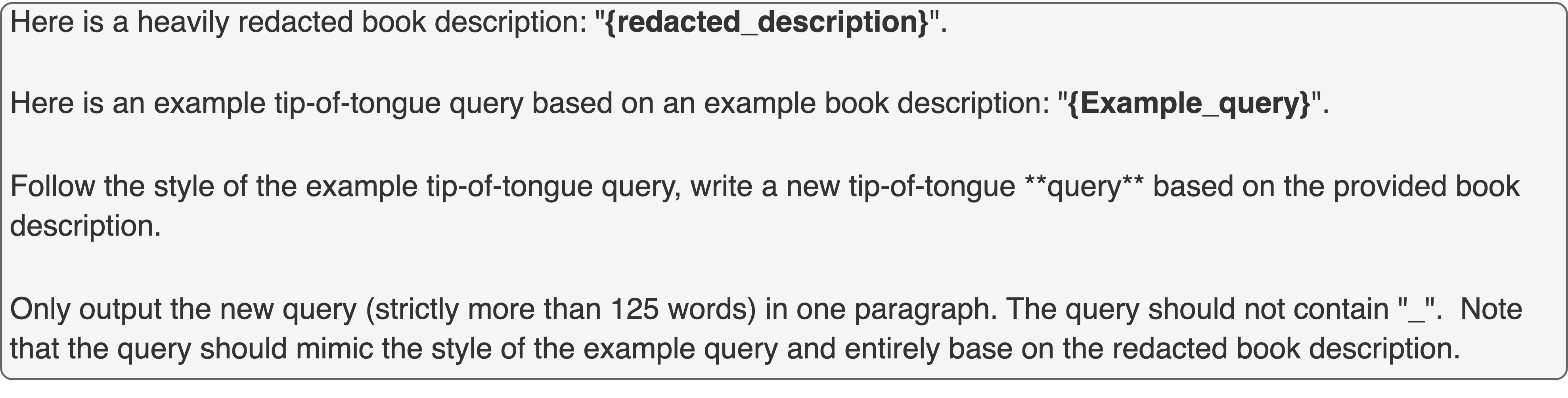}
    \caption{WhatsThatBook Document regularization 60\%, 80\% Prompt}
    \label{fig: wtb_d_reg_60_80_prompt}
\end{figure*}

\begin{figure*}[ht]
    \centering
        \includegraphics[width=\linewidth]{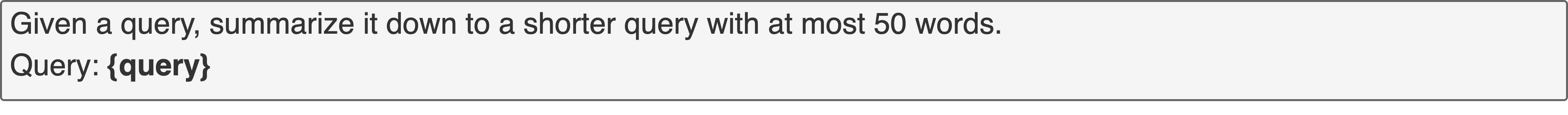}
    \caption{WhatsThatBook Query Regularization on synthetic queries from Doc Reg 40\%, 60\%, and 80\%}
    \label{fig: wtb_q_reg_d_reg_prompt}
\end{figure*}

\begin{figure*}[ht]
    \centering
        \includegraphics[width=\linewidth]{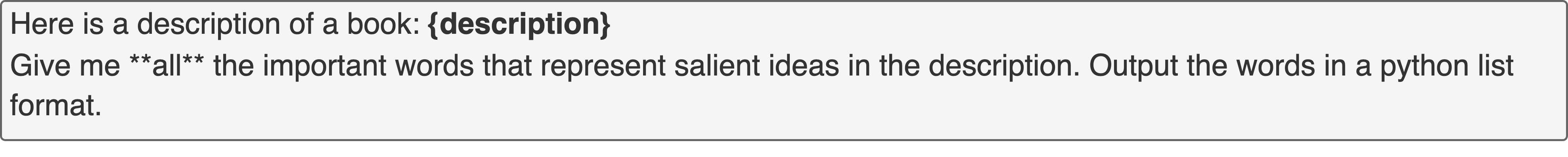}
    \caption{WhatsThatBook Keywords Extraction Prompt}
    \label{fig: wtb_keyword_prompt}
\end{figure*}

\begin{figure*}[ht]
    \centering
        \includegraphics[width=\linewidth]{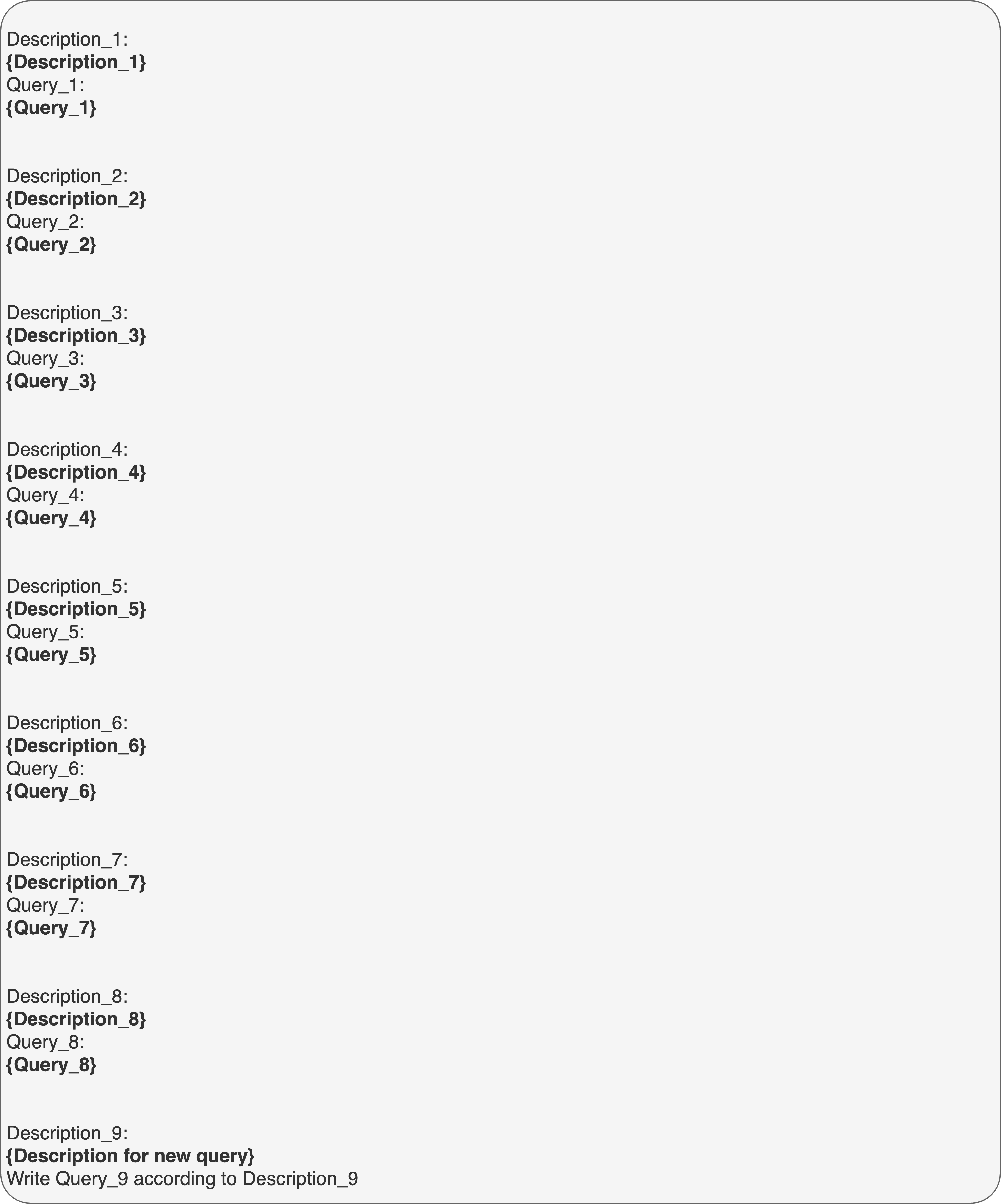}
    \caption{WhatsThatBook Promptagator Prompt}
    \label{fig: wtb_promptagator_prompt}
\end{figure*}

\begin{itemize}
    \item Figure \ref{fig: wtb_I_reg_prompt} for Instruction Regularization prompt
    \item Figure \ref{fig: wtb_q_reg_I_reg_prompt} for Query Regularization on Instruction Regularization prompt
    \item Figure \ref{fig: wtb_d_reg_40_prompt} for Document Regularization 40\% prompt
    \item Figure \ref{fig: wtb_d_reg_60_80_prompt} for Document Regularization 60\%, 80\% prompt
    \item Figure \ref{fig: wtb_q_reg_d_reg_prompt} for Query Regularization on synthetic queries from Doc Reg 40\%, 60\%, and 80\%
    \item Figure \ref{fig: wtb_keyword_prompt} for keywords extraction prompt
    \item Figure \ref{fig: wtb_promptagator_prompt} for Promptagator prompt example
\end{itemize}

\section{Experiment Details}
\label{sec: appendix_experiment_details}
We provide the experimental results on the three datasets in Table \ref{tab: doris-mae full table}, \ref{tab: ArguAna full experiment result}, \ref{tab: wtb full experiment result} with a more complete set of metrics. The results are consistent with those reported in Section \ref{sec: results}.

\begin{table*}[!ht]
\centering
\scriptsize
\begin{tabular}{llllllll}
\toprule
Model (Method) & R@5 & R@10 & R@20 & RP & NDCG@10 & MRR@10 & MAP \\
\midrule
\multicolumn{8}{c}{E5-Large-v2} \\
\midrule

Pretrained  & 14.67 & 25.98 & 42.15 & 38.18 & 71.98 & 14.34 & 40.52\\

Contrastive Fine-tuning  & 15.28 & 25.97 & 39.87 & 34.73 & 72.51 & 13.43 & 38.87\\
\cmidrule(lr){1-1}

Promptagator  & 16.18 & 26.58 & 45.59 & 38.99 & 73.95 & 14.83 & 42.69\\
\cmidrule(lr){1-1}

Document\textsubscript{40\% reg}  & 14.90 \color{red} $\downarrow$ -1.28  & 27.65 \color{darkgreen} $\uparrow$ 1.07  & 46.55 \color{darkgreen} $\uparrow$ 0.96  & 39.33 & 74.28 \color{darkgreen} $\uparrow$ 0.33  & 14.87 & 43.10 \color{darkgreen} $\uparrow$ 0.41 \\

Query\textsubscript{reg} $\circ$  Document\textsubscript{40\% reg}  & 15.85 & 27.44 \color{darkgreen} $\uparrow$ 0.86  & 45.67 & 39.73 \color{darkgreen} $\uparrow$ 0.74  & 73.96 & 16.10 \color{darkgreen} $\uparrow$ 1.27  & 43.06 \color{darkgreen} $\uparrow$ 0.37 \\

Document\textsubscript{60\% reg}  & 15.23 \color{red} $\downarrow$ -0.95  & 27.68 \color{darkgreen} $\uparrow$ 1.10  & 46.57 \color{darkgreen} $\uparrow$ 0.98  & 38.86 & 74.98 \color{darkgreen} $\uparrow$ 1.03  & 15.14 & 43.26 \color{darkgreen} $\uparrow$ 0.57 \\

Query\textsubscript{reg} $\circ$  Document\textsubscript{60\% reg}  & 15.31 \color{red} $\downarrow$ -0.87  & 27.74 \color{darkgreen} $\uparrow$ 1.16  & 48.16 \color{darkgreen} $\uparrow$ 2.57  & 39.60 \color{darkgreen} $\uparrow$ 0.61  & 74.90 \color{darkgreen} $\uparrow$ 0.95  & 15.43 \color{darkgreen} $\uparrow$ 0.60  & 44.48 \color{darkgreen} $\uparrow$ 1.79 \\

Document\textsubscript{80\% reg}  & 14.68 \color{red} $\downarrow$ -1.50  & 25.02 \color{red} $\downarrow$ -1.56  & 43.89 \color{red} $\downarrow$ -1.70  & 38.91 & 74.13 \color{darkgreen} $\uparrow$ 0.18  & 13.51 \color{red} $\downarrow$ -1.32  & 43.31 \color{darkgreen} $\uparrow$ 0.62 \\

Query\textsubscript{reg} $\circ$  Document\textsubscript{80\% reg}  & 15.63 \color{red} $\downarrow$ -0.55  & 27.93 \color{darkgreen} $\uparrow$ 1.35  & 46.75 \color{darkgreen} $\uparrow$ 1.16  & 39.37 & 75.08 \color{darkgreen} $\uparrow$ 1.13  & 16.62 \color{darkgreen} $\uparrow$ 1.79  & 44.24 \color{darkgreen} $\uparrow$ 1.55 \\

Instr\textsubscript{reg}  & \textbf{16.38} & 26.26 & 46.73 \color{darkgreen} $\uparrow$ 1.14  & 40.31 \color{darkgreen} $\uparrow$ 1.32  & 74.42 \color{darkgreen} $\uparrow$ 0.47  & 17.06 \color{darkgreen} $\uparrow$ 2.23  & 44.84 \color{darkgreen} $\uparrow$ 2.15 \\

Query\textsubscript{reg} $\circ$  Instr\textsubscript{reg}  & 15.76 \color{red} $\downarrow$ -0.42  & \textbf{30.56} \color{darkgreen} $\uparrow$ 3.98  & \textbf{48.46} \color{darkgreen} $\uparrow$ 2.87  & \textbf{42.25} \color{darkgreen} $\uparrow$ 3.26  & \textbf{76.02} \color{darkgreen} $\uparrow$ 2.07  & \textbf{20.86} \color{darkgreen} $\uparrow$ 6.03  & \textbf{46.29} \color{darkgreen} $\uparrow$ 3.60 \\
\midrule
\multicolumn{8}{c}{SimCSE} \\
\midrule

Pretrained  & 14.38 & 23.24 & 41.83 & 36.27 & 70.81 & \textbf{24.83} & 42.02\\

Contrastive Fine-tuning  & 15.60 & 27.70 & 44.99 & 39.06 & 73.18 & 22.41 & 42.87\\
\cmidrule(lr){1-1}

Promptagator  & 16.35 & 28.98 & 46.69 & 40.64 & 73.97 & 16.54 & 44.12\\
\cmidrule(lr){1-1}

Document\textsubscript{40\% reg}  & \textbf{17.70} \color{darkgreen} $\uparrow$ 1.35  & \textbf{29.85} \color{darkgreen} $\uparrow$ 0.87  & 45.30 \color{red} $\downarrow$ -1.39  & 40.43 & 74.49 \color{darkgreen} $\uparrow$ 0.52  & 16.49 & 44.67 \color{darkgreen} $\uparrow$ 0.55 \\

Query\textsubscript{reg} $\circ$  Document\textsubscript{40\% reg}  & 16.06 & 28.77 & 45.46 \color{red} $\downarrow$ -1.23  & 40.41 & 73.23 \color{red} $\downarrow$ -0.74  & 16.56 & 43.70 \color{red} $\downarrow$ -0.42 \\

Document\textsubscript{60\% reg}  & 17.08 \color{darkgreen} $\uparrow$ 0.73  & 28.80 & 45.33 \color{red} $\downarrow$ -1.36  & 39.91 \color{red} $\downarrow$ -0.73  & 74.70 \color{darkgreen} $\uparrow$ 0.73  & 17.75 \color{darkgreen} $\uparrow$ 1.21  & 44.50 \color{darkgreen} $\uparrow$ 0.38 \\

Query\textsubscript{reg} $\circ$  Document\textsubscript{60\% reg}  & 17.04 \color{darkgreen} $\uparrow$ 0.69  & 28.63 & 46.68 & 40.59 & 74.65 \color{darkgreen} $\uparrow$ 0.68  & 18.32 \color{darkgreen} $\uparrow$ 1.78  & 45.05 \color{darkgreen} $\uparrow$ 0.93 \\

Document\textsubscript{80\% reg}  & 15.01 \color{red} $\downarrow$ -1.34  & 26.37 \color{red} $\downarrow$ -2.61  & 42.53 \color{red} $\downarrow$ -4.16  & 38.32 \color{red} $\downarrow$ -2.32  & 73.69 \color{red} $\downarrow$ -0.28  & 18.68 \color{darkgreen} $\uparrow$ 2.14  & 42.22 \color{red} $\downarrow$ -1.90 \\

Query\textsubscript{reg} $\circ$  Document\textsubscript{80\% reg}  & 15.92 & 27.32 \color{red} $\downarrow$ -1.66  & 45.70 \color{red} $\downarrow$ -0.99  & 39.59 \color{red} $\downarrow$ -1.05  & 73.86 & 18.54 \color{darkgreen} $\uparrow$ 2.00  & 43.96\\

Instr\textsubscript{reg}  & 16.50 & 27.81 \color{red} $\downarrow$ -1.17  & 46.68 & \textbf{40.71} & \textbf{75.60} \color{darkgreen} $\uparrow$ 1.63  & 19.71 \color{darkgreen} $\uparrow$ 3.17  & \textbf{45.85} \color{darkgreen} $\uparrow$ 1.73 \\

Query\textsubscript{reg} $\circ$  Instr\textsubscript{reg}  & 16.06 & 28.25 \color{red} $\downarrow$ -0.73  & \textbf{47.61} \color{darkgreen} $\uparrow$ 0.92  & 39.88 \color{red} $\downarrow$ -0.76  & 75.43 \color{darkgreen} $\uparrow$ 1.46  & 22.14 \color{darkgreen} $\uparrow$ 5.60  & 45.52 \color{darkgreen} $\uparrow$ 1.40 \\
\midrule
\multicolumn{8}{c}{RoBERTa} \\
\midrule

Pretrained  & 11.96 & 21.45 & 34.77 & 30.68 & 66.86 & 8.56 & 34.14\\

Contrastive Fine-tuning  & 14.75 & 25.95 & 43.32 & 37.68 & 73.55 & \textbf{20.79} & 41.67\\
\cmidrule(lr){1-1}

Promptagator  & 13.03 & 24.50 & 42.88 & 36.65 & 72.23 & 16.82 & 40.19\\
\cmidrule(lr){1-1}

Document\textsubscript{40\% reg}  & 12.54 & 24.94 & 44.89 \color{darkgreen} $\uparrow$ 2.01  & 36.86 & 72.94 \color{darkgreen} $\uparrow$ 0.71  & 18.16 \color{darkgreen} $\uparrow$ 1.34  & 41.65 \color{darkgreen} $\uparrow$ 1.46 \\

Query\textsubscript{reg} $\circ$  Document\textsubscript{40\% reg}  & 14.66 \color{darkgreen} $\uparrow$ 1.63  & 27.58 \color{darkgreen} $\uparrow$ 3.08  & 47.41 \color{darkgreen} $\uparrow$ 4.53  & 38.54 \color{darkgreen} $\uparrow$ 1.89  & 74.04 \color{darkgreen} $\uparrow$ 1.81  & 18.81 \color{darkgreen} $\uparrow$ 1.99  & 42.45 \color{darkgreen} $\uparrow$ 2.26 \\

Document\textsubscript{60\% reg}  & 14.11 \color{darkgreen} $\uparrow$ 1.08  & 25.80 \color{darkgreen} $\uparrow$ 1.30  & 43.99 \color{darkgreen} $\uparrow$ 1.11  & 37.03 & 72.94 \color{darkgreen} $\uparrow$ 0.71  & 16.84 & 41.77 \color{darkgreen} $\uparrow$ 1.58 \\

Query\textsubscript{reg} $\circ$  Document\textsubscript{60\% reg}  & 14.75 \color{darkgreen} $\uparrow$ 1.72  & 28.34 \color{darkgreen} $\uparrow$ 3.84  & 47.25 \color{darkgreen} $\uparrow$ 4.37  & 39.43 \color{darkgreen} $\uparrow$ 2.78  & 74.61 \color{darkgreen} $\uparrow$ 2.38  & 20.48 \color{darkgreen} $\uparrow$ 3.66  & 43.33 \color{darkgreen} $\uparrow$ 3.14 \\

Document\textsubscript{80\% reg}  & 14.47 \color{darkgreen} $\uparrow$ 1.44  & 25.53 \color{darkgreen} $\uparrow$ 1.03  & 42.47 & 37.57 \color{darkgreen} $\uparrow$ 0.92  & 72.61 \color{darkgreen} $\uparrow$ 0.38  & 15.10 \color{red} $\downarrow$ -1.72  & 41.55 \color{darkgreen} $\uparrow$ 1.36 \\

Query\textsubscript{reg} $\circ$  Document\textsubscript{80\% reg}  & \textbf{15.84} \color{darkgreen} $\uparrow$ 2.81  & \textbf{27.96} \color{darkgreen} $\uparrow$ 3.46  & 46.96 \color{darkgreen} $\uparrow$ 4.08  & 39.09 \color{darkgreen} $\uparrow$ 2.44  & \textbf{74.71} \color{darkgreen} $\uparrow$ 2.48  & 20.49 \color{darkgreen} $\uparrow$ 3.67  & \textbf{43.41} \color{darkgreen} $\uparrow$ 3.22 \\

Instr\textsubscript{reg}  & 13.80 \color{darkgreen} $\uparrow$ 0.77  & 25.00 & 45.78 \color{darkgreen} $\uparrow$ 2.90  & 36.29 & 73.53 \color{darkgreen} $\uparrow$ 1.30  & 18.24 \color{darkgreen} $\uparrow$ 1.42  & 41.87 \color{darkgreen} $\uparrow$ 1.68 \\

Query\textsubscript{reg} $\circ$  Instr\textsubscript{reg}  & 15.20 \color{darkgreen} $\uparrow$ 2.17  & 26.88 \color{darkgreen} $\uparrow$ 2.38  & \textbf{47.44} \color{darkgreen} $\uparrow$ 4.56  & \textbf{39.38} \color{darkgreen} $\uparrow$ 2.73  & 74.08 \color{darkgreen} $\uparrow$ 1.85  & 19.31 \color{darkgreen} $\uparrow$ 2.49  & 42.52 \color{darkgreen} $\uparrow$ 2.33 \\
\midrule
\multicolumn{8}{c}{SPECTER-v2} \\
\midrule

Pretrained  & 13.59 & 25.63 & 41.92 & 35.91 & 71.46 & 21.74 & 38.84\\

Contrastive Fine-tuning  & 14.56 & 25.18 & \textbf{45.64} & 36.76 & 71.46 & 20.49 & 40.45\\
\cmidrule(lr){1-1}

Promptagator  & 14.13 & 23.67 & 39.22 & 36.70 & 71.55 & 22.37 & 39.10\\
\cmidrule(lr){1-1}

Document\textsubscript{40\% reg}  & 15.39 \color{darkgreen} $\uparrow$ 1.26  & 25.15 \color{darkgreen} $\uparrow$ 1.48  & 43.39 \color{darkgreen} $\uparrow$ 4.17  & 36.89 & 72.11 \color{darkgreen} $\uparrow$ 0.56  & 21.66 \color{red} $\downarrow$ -0.71  & 40.48 \color{darkgreen} $\uparrow$ 1.38 \\

Query\textsubscript{reg} $\circ$  Document\textsubscript{40\% reg}  & 14.19 & 24.99 \color{darkgreen} $\uparrow$ 1.32  & 42.66 \color{darkgreen} $\uparrow$ 3.44  & 36.39 & 71.49 & 21.40 \color{red} $\downarrow$ -0.97  & 39.77 \color{darkgreen} $\uparrow$ 0.67 \\

Document\textsubscript{60\% reg}  & 15.41 \color{darkgreen} $\uparrow$ 1.28  & 26.79 \color{darkgreen} $\uparrow$ 3.12  & 44.19 \color{darkgreen} $\uparrow$ 4.97  & 36.90 & 72.81 \color{darkgreen} $\uparrow$ 1.26  & 21.35 \color{red} $\downarrow$ -1.02  & 41.07 \color{darkgreen} $\uparrow$ 1.97 \\

Query\textsubscript{reg} $\circ$  Document\textsubscript{60\% reg}  & 14.95 \color{darkgreen} $\uparrow$ 0.82  & 25.81 \color{darkgreen} $\uparrow$ 2.14  & 43.30 \color{darkgreen} $\uparrow$ 4.08  & 38.38 \color{darkgreen} $\uparrow$ 1.68  & 72.19 \color{darkgreen} $\uparrow$ 0.64  & 21.60 \color{red} $\downarrow$ -0.77  & 40.55 \color{darkgreen} $\uparrow$ 1.45 \\

Document\textsubscript{80\% reg}  & \textbf{17.02} \color{darkgreen} $\uparrow$ 2.89  & \textbf{28.63} \color{darkgreen} $\uparrow$ 4.96  & 44.97 \color{darkgreen} $\uparrow$ 5.75  & \textbf{38.35} \color{darkgreen} $\uparrow$ 1.65  & \textbf{73.94} \color{darkgreen} $\uparrow$ 2.39  & \textbf{22.78} \color{darkgreen} $\uparrow$ 0.41  & \textbf{42.37} \color{darkgreen} $\uparrow$ 3.27 \\

Query\textsubscript{reg} $\circ$  Document\textsubscript{80\% reg}  & 14.70 \color{darkgreen} $\uparrow$ 0.57  & 26.85 \color{darkgreen} $\uparrow$ 3.18  & 44.66 \color{darkgreen} $\uparrow$ 5.44  & 37.61 \color{darkgreen} $\uparrow$ 0.91  & 73.07 \color{darkgreen} $\uparrow$ 1.52  & 22.65 & 41.70 \color{darkgreen} $\uparrow$ 2.60 \\

Instr\textsubscript{reg}  & 14.73 \color{darkgreen} $\uparrow$ 0.60  & 25.82 \color{darkgreen} $\uparrow$ 2.15  & 41.50 \color{darkgreen} $\uparrow$ 2.28  & 35.91 \color{red} $\downarrow$ -0.79  & 73.06 \color{darkgreen} $\uparrow$ 1.51  & 21.05 \color{red} $\downarrow$ -1.32  & 39.24 \color{darkgreen} $\uparrow$ 0.14 \\

Query\textsubscript{reg} $\circ$  Instr\textsubscript{reg}  & 14.50 \color{darkgreen} $\uparrow$ 0.37  & 25.32 \color{darkgreen} $\uparrow$ 1.65  & 41.44 \color{darkgreen} $\uparrow$ 2.22  & 36.01 \color{red} $\downarrow$ -0.69  & 72.46 \color{darkgreen} $\uparrow$ 0.91  & 22.26 & 39.19\\

\bottomrule
\end{tabular}
\caption{DORIS-MAE Full Experimental Results}
\label{tab: doris-mae full table}
\end{table*}

\begin{table*}[!ht]
\centering
\scriptsize
\begin{tabular}{llllllll}
\toprule
Model (Method) & R@5 & R@10 & R@20 & RP & NDCG@10 & MRR@10 & MAP \\
\midrule
\multicolumn{8}{c}{E5-Large-v2} \\
\midrule

Pretrained  & 59.37 & 76.74 & 88.16 & 23.33 & 47.75 & 38.96 & 39.84\\

Contrastive Fine-tuning  & 70.69 & 85.86 & 94.77 & 29.34 & 56.15 & 47.13 & 47.57\\
\cmidrule(lr){1-1}

Promptagator  & 69.41 & 83.35 & 94.11 & 30.38 & 55.14 & 46.67 & 47.24\\
\cmidrule(lr){1-1}

Document\textsubscript{40\% reg}  & \textbf{78.72} \color{darkgreen} $\uparrow$ 9.31  & 90.43 \color{darkgreen} $\uparrow$ 7.08  & \textbf{96.46} \color{darkgreen} $\uparrow$ 2.35  & 39.40 \color{darkgreen} $\uparrow$ 9.02  & 64.24 \color{darkgreen} $\uparrow$ 9.10  & 56.45 \color{darkgreen} $\uparrow$ 9.78  & 56.46 \color{darkgreen} $\uparrow$ 9.22 \\

Query\textsubscript{reg} $\circ$  Document\textsubscript{40\% reg}  & 77.76 \color{darkgreen} $\uparrow$ 8.35  & 88.89 \color{darkgreen} $\uparrow$ 5.54  & 96.21 \color{darkgreen} $\uparrow$ 2.10  & 38.20 \color{darkgreen} $\uparrow$ 7.82  & 63.00 \color{darkgreen} $\uparrow$ 7.86  & 55.24 \color{darkgreen} $\uparrow$ 8.57  & 55.38 \color{darkgreen} $\uparrow$ 8.14 \\

Document\textsubscript{60\% reg}  & 78.16 \color{darkgreen} $\uparrow$ 8.75  & \textbf{90.85} \color{darkgreen} $\uparrow$ 7.50  & 96.41 \color{darkgreen} $\uparrow$ 2.30  & 39.63 \color{darkgreen} $\uparrow$ 9.25  & \textbf{64.41} \color{darkgreen} $\uparrow$ 9.27  & \textbf{56.57} \color{darkgreen} $\uparrow$ 9.90  & \textbf{56.53} \color{darkgreen} $\uparrow$ 9.29 \\

Query\textsubscript{reg} $\circ$  Document\textsubscript{60\% reg}  & 78.34 \color{darkgreen} $\uparrow$ 8.93  & 90.52 \color{darkgreen} $\uparrow$ 7.17  & 96.37 \color{darkgreen} $\uparrow$ 2.26  & 39.06 \color{darkgreen} $\uparrow$ 8.68  & 64.10 \color{darkgreen} $\uparrow$ 8.96  & 56.24 \color{darkgreen} $\uparrow$ 9.57  & 56.24 \color{darkgreen} $\uparrow$ 9.00 \\

Document\textsubscript{80\% reg}  & 77.37 \color{darkgreen} $\uparrow$ 7.96  & 89.81 \color{darkgreen} $\uparrow$ 6.46  & 95.57 \color{darkgreen} $\uparrow$ 1.46  & 38.14 \color{darkgreen} $\uparrow$ 7.76  & 63.14 \color{darkgreen} $\uparrow$ 8.00  & 55.16 \color{darkgreen} $\uparrow$ 8.49  & 55.21 \color{darkgreen} $\uparrow$ 7.97 \\

Query\textsubscript{reg} $\circ$  Document\textsubscript{80\% reg}  & 78.44 \color{darkgreen} $\uparrow$ 9.03  & 90.82 \color{darkgreen} $\uparrow$ 7.47  & 96.25 \color{darkgreen} $\uparrow$ 2.14  & \textbf{39.69} \color{darkgreen} $\uparrow$ 9.31  & 64.36 \color{darkgreen} $\uparrow$ 9.22  & 56.49 \color{darkgreen} $\uparrow$ 9.82  & 56.45 \color{darkgreen} $\uparrow$ 9.21 \\

Instr\textsubscript{reg}  & 72.06 \color{darkgreen} $\uparrow$ 2.65  & 86.23 \color{darkgreen} $\uparrow$ 2.88  & 95.70 \color{darkgreen} $\uparrow$ 1.59  & 32.14 \color{darkgreen} $\uparrow$ 1.76  & 57.76 \color{darkgreen} $\uparrow$ 2.62  & 49.17 \color{darkgreen} $\uparrow$ 2.50  & 49.58 \color{darkgreen} $\uparrow$ 2.34 \\

Query\textsubscript{reg} $\circ$  Instr\textsubscript{reg}  & 71.59 \color{darkgreen} $\uparrow$ 2.18  & 85.12 \color{darkgreen} $\uparrow$ 1.77  & 95.18 \color{darkgreen} $\uparrow$ 1.07  & 31.65 \color{darkgreen} $\uparrow$ 1.27  & 56.89 \color{darkgreen} $\uparrow$ 1.75  & 48.37 \color{darkgreen} $\uparrow$ 1.70  & 48.86 \color{darkgreen} $\uparrow$ 1.62 \\
\midrule
\multicolumn{8}{c}{SimCSE} \\
\midrule

Pretrained  & 49.03 & 68.92 & 81.12 & 15.65 & 39.23 & 30.35 & 31.42\\

Contrastive Fine-tuning  & 36.70 & 51.22 & 64.79 & 11.18 & 28.85 & 22.12 & 23.57\\
\cmidrule(lr){1-1}

Promptagator  & 69.49 & 84.45 & 92.54 & 31.91 & 56.38 & 47.97 & 48.36\\
\cmidrule(lr){1-1}

Document\textsubscript{40\% reg}  & 74.92 \color{darkgreen} $\uparrow$ 5.43  & 86.37 \color{darkgreen} $\uparrow$ 1.92  & 93.36 \color{darkgreen} $\uparrow$ 0.82  & 35.54 \color{darkgreen} $\uparrow$ 3.63  & 60.24 \color{darkgreen} $\uparrow$ 3.86  & 52.42 \color{darkgreen} $\uparrow$ 4.45  & 52.60 \color{darkgreen} $\uparrow$ 4.24 \\

Query\textsubscript{reg} $\circ$  Document\textsubscript{40\% reg}  & \textbf{76.07} \color{darkgreen} $\uparrow$ 6.58  & \textbf{87.46} \color{darkgreen} $\uparrow$ 3.01  & \textbf{93.97} \color{darkgreen} $\uparrow$ 1.43  & \textbf{36.96} \color{darkgreen} $\uparrow$ 5.05  & \textbf{61.38} \color{darkgreen} $\uparrow$ 5.00  & \textbf{53.57} \color{darkgreen} $\uparrow$ 5.60  & \textbf{53.73} \color{darkgreen} $\uparrow$ 5.37 \\

Document\textsubscript{60\% reg}  & 73.20 \color{darkgreen} $\uparrow$ 3.71  & 84.71 & 92.10 & 33.47 \color{darkgreen} $\uparrow$ 1.56  & 58.46 \color{darkgreen} $\uparrow$ 2.08  & 50.58 \color{darkgreen} $\uparrow$ 2.61  & 50.84 \color{darkgreen} $\uparrow$ 2.48 \\

Query\textsubscript{reg} $\circ$  Document\textsubscript{60\% reg}  & 75.75 \color{darkgreen} $\uparrow$ 6.26  & 86.58 \color{darkgreen} $\uparrow$ 2.13  & 93.19 \color{darkgreen} $\uparrow$ 0.65  & 36.68 \color{darkgreen} $\uparrow$ 4.77  & 60.94 \color{darkgreen} $\uparrow$ 4.56  & 53.24 \color{darkgreen} $\uparrow$ 5.27  & 53.44 \color{darkgreen} $\uparrow$ 5.08 \\

Document\textsubscript{80\% reg}  & 70.38 & 81.60 \color{red} $\downarrow$ -2.85  & 89.67 \color{red} $\downarrow$ -2.87  & 31.42 & 55.76 & 47.97 & 48.39\\

Query\textsubscript{reg} $\circ$  Document\textsubscript{80\% reg}  & 74.10 \color{darkgreen} $\uparrow$ 4.61  & 85.13 & 92.59 & 35.71 \color{darkgreen} $\uparrow$ 3.80  & 59.79 \color{darkgreen} $\uparrow$ 3.41  & 52.18 \color{darkgreen} $\uparrow$ 4.21  & 52.45 \color{darkgreen} $\uparrow$ 4.09 \\

Instr\textsubscript{reg}  & 70.97 & 86.14 \color{darkgreen} $\uparrow$ 1.69  & 93.91 \color{darkgreen} $\uparrow$ 1.37  & 32.18 & 57.41 & 48.80 & 49.14\\

Query\textsubscript{reg} $\circ$  Instr\textsubscript{reg}  & 71.46 \color{darkgreen} $\uparrow$ 1.97  & 86.34 \color{darkgreen} $\uparrow$ 1.89  & 93.78 \color{darkgreen} $\uparrow$ 1.24  & 33.18 \color{darkgreen} $\uparrow$ 1.27  & 58.16 \color{darkgreen} $\uparrow$ 1.78  & 49.78 \color{darkgreen} $\uparrow$ 1.81  & 50.04 \color{darkgreen} $\uparrow$ 1.68 \\
\midrule
\multicolumn{8}{c}{RoBERTa} \\
\midrule

Pretrained  & 30.51 & 40.56 & 49.68 & 9.84 & 23.56 & 18.49 & 19.54\\

Contrastive Fine-tuning  & 59.12 & 74.36 & 84.67 & 24.33 & 47.57 & 39.65 & 40.29\\
\cmidrule(lr){1-1}

Promptagator  & 61.51 & 76.67 & 88.19 & 26.16 & 49.80 & 41.85 & 42.51\\
\cmidrule(lr){1-1}

Document\textsubscript{40\% reg}  & 67.38 \color{darkgreen} $\uparrow$ 5.87  & \textbf{81.87} \color{darkgreen} $\uparrow$ 5.20  & 91.03 \color{darkgreen} $\uparrow$ 2.84  & 30.26 \color{darkgreen} $\uparrow$ 4.10  & 54.54 \color{darkgreen} $\uparrow$ 4.74  & 46.43 \color{darkgreen} $\uparrow$ 4.58  & 46.85 \color{darkgreen} $\uparrow$ 4.34 \\

Query\textsubscript{reg} $\circ$  Document\textsubscript{40\% reg}  & 67.29 \color{darkgreen} $\uparrow$ 5.78  & 81.64 \color{darkgreen} $\uparrow$ 4.97  & \textbf{91.08} \color{darkgreen} $\uparrow$ 2.89  & 29.21 \color{darkgreen} $\uparrow$ 3.05  & 54.04 \color{darkgreen} $\uparrow$ 4.24  & 45.85 \color{darkgreen} $\uparrow$ 4.00  & 46.27 \color{darkgreen} $\uparrow$ 3.76 \\

Document\textsubscript{60\% reg}  & 67.51 \color{darkgreen} $\uparrow$ 6.00  & 81.78 \color{darkgreen} $\uparrow$ 5.11  & 90.77 \color{darkgreen} $\uparrow$ 2.58  & \textbf{30.50} \color{darkgreen} $\uparrow$ 4.34  & \textbf{54.85} \color{darkgreen} $\uparrow$ 5.05  & \textbf{46.85} \color{darkgreen} $\uparrow$ 5.00  & \textbf{47.26} \color{darkgreen} $\uparrow$ 4.75 \\

Query\textsubscript{reg} $\circ$  Document\textsubscript{60\% reg}  & 67.71 \color{darkgreen} $\uparrow$ 6.20  & 81.71 \color{darkgreen} $\uparrow$ 5.04  & 90.99 \color{darkgreen} $\uparrow$ 2.80  & 29.74 \color{darkgreen} $\uparrow$ 3.58  & 54.41 \color{darkgreen} $\uparrow$ 4.61  & 46.34 \color{darkgreen} $\uparrow$ 4.49  & 46.73 \color{darkgreen} $\uparrow$ 4.22 \\

Document\textsubscript{80\% reg}  & \textbf{67.82} \color{darkgreen} $\uparrow$ 6.31  & 80.97 \color{darkgreen} $\uparrow$ 4.30  & 90.18 \color{darkgreen} $\uparrow$ 1.99  & 30.47 \color{darkgreen} $\uparrow$ 4.31  & 54.62 \color{darkgreen} $\uparrow$ 4.82  & 46.79 \color{darkgreen} $\uparrow$ 4.94  & 47.20 \color{darkgreen} $\uparrow$ 4.69 \\

Query\textsubscript{reg} $\circ$  Document\textsubscript{80\% reg}  & 67.38 \color{darkgreen} $\uparrow$ 5.87  & 81.42 \color{darkgreen} $\uparrow$ 4.75  & 90.70 \color{darkgreen} $\uparrow$ 2.51  & 30.09 \color{darkgreen} $\uparrow$ 3.93  & 54.53 \color{darkgreen} $\uparrow$ 4.73  & 46.59 \color{darkgreen} $\uparrow$ 4.74  & 46.97 \color{darkgreen} $\uparrow$ 4.46 \\

Instr\textsubscript{reg}  & 64.47 \color{darkgreen} $\uparrow$ 2.96  & 79.29 \color{darkgreen} $\uparrow$ 2.62  & 90.09 \color{darkgreen} $\uparrow$ 1.90  & 26.77 \color{darkgreen} $\uparrow$ 0.61  & 51.49 \color{darkgreen} $\uparrow$ 1.69  & 43.20 \color{darkgreen} $\uparrow$ 1.35  & 43.78 \color{darkgreen} $\uparrow$ 1.27 \\

Query\textsubscript{reg} $\circ$  Instr\textsubscript{reg}  & 65.14 \color{darkgreen} $\uparrow$ 3.63  & 80.38 \color{darkgreen} $\uparrow$ 3.71  & 90.26 \color{darkgreen} $\uparrow$ 2.07  & 27.59 \color{darkgreen} $\uparrow$ 1.43  & 52.25 \color{darkgreen} $\uparrow$ 2.45  & 43.89 \color{darkgreen} $\uparrow$ 2.04  & 44.39 \color{darkgreen} $\uparrow$ 1.88 \\
\midrule
\multicolumn{8}{c}{SPECTER-v2} \\
\midrule

Pretrained  & 38.91 & 56.57 & 71.79 & 11.06 & 30.91 & 23.23 & 24.65\\

Contrastive Fine-tuning  & 41.90 & 59.65 & 75.70 & 12.25 & 33.20 & 25.26 & 26.75\\
\cmidrule(lr){1-1}

Promptagator  & 38.64 & 58.34 & 75.15 & 9.59 & 30.24 & 21.77 & 23.40\\
\cmidrule(lr){1-1}

Document\textsubscript{40\% reg}  & 56.09 \color{darkgreen} $\uparrow$ 17.45  & 74.12 \color{darkgreen} $\uparrow$ 15.78  & 86.32 \color{darkgreen} $\uparrow$ 11.17  & 21.02 \color{darkgreen} $\uparrow$ 11.43  & 45.00 \color{darkgreen} $\uparrow$ 14.76  & 36.30 \color{darkgreen} $\uparrow$ 14.53  & 37.18 \color{darkgreen} $\uparrow$ 13.78 \\

Query\textsubscript{reg} $\circ$  Document\textsubscript{40\% reg}  & 51.28 \color{darkgreen} $\uparrow$ 12.64  & 69.75 \color{darkgreen} $\uparrow$ 11.41  & 83.33 \color{darkgreen} $\uparrow$ 8.18  & 17.57 \color{darkgreen} $\uparrow$ 7.98  & 40.59 \color{darkgreen} $\uparrow$ 10.35  & 31.81 \color{darkgreen} $\uparrow$ 10.04  & 32.96 \color{darkgreen} $\uparrow$ 9.56 \\

Document\textsubscript{60\% reg}  & 58.87 \color{darkgreen} $\uparrow$ 20.23  & 75.45 \color{darkgreen} $\uparrow$ 17.11  & 86.93 \color{darkgreen} $\uparrow$ 11.78  & 22.11 \color{darkgreen} $\uparrow$ 12.52  & 46.55 \color{darkgreen} $\uparrow$ 16.31  & 37.88 \color{darkgreen} $\uparrow$ 16.11  & 38.69 \color{darkgreen} $\uparrow$ 15.29 \\

Query\textsubscript{reg} $\circ$  Document\textsubscript{60\% reg}  & 54.22 \color{darkgreen} $\uparrow$ 15.58  & 71.81 \color{darkgreen} $\uparrow$ 13.47  & 84.97 \color{darkgreen} $\uparrow$ 9.82  & 19.35 \color{darkgreen} $\uparrow$ 9.76  & 42.90 \color{darkgreen} $\uparrow$ 12.66  & 34.21 \color{darkgreen} $\uparrow$ 12.44  & 35.26 \color{darkgreen} $\uparrow$ 11.86 \\

Document\textsubscript{80\% reg}  & \textbf{61.02} \color{darkgreen} $\uparrow$ 22.38  & \textbf{76.84} \color{darkgreen} $\uparrow$ 18.50  & \textbf{87.48} \color{darkgreen} $\uparrow$ 12.33  & \textbf{24.39} \color{darkgreen} $\uparrow$ 14.80  & \textbf{48.68} \color{darkgreen} $\uparrow$ 18.44  & \textbf{40.24} \color{darkgreen} $\uparrow$ 18.47  & \textbf{40.94} \color{darkgreen} $\uparrow$ 17.54 \\

Query\textsubscript{reg} $\circ$  Document\textsubscript{80\% reg}  & 57.55 \color{darkgreen} $\uparrow$ 18.91  & 74.67 \color{darkgreen} $\uparrow$ 16.33  & 86.65 \color{darkgreen} $\uparrow$ 11.50  & 22.00 \color{darkgreen} $\uparrow$ 12.41  & 46.01 \color{darkgreen} $\uparrow$ 15.77  & 37.44 \color{darkgreen} $\uparrow$ 15.67  & 38.27 \color{darkgreen} $\uparrow$ 14.87 \\

Instr\textsubscript{reg}  & 43.26 \color{darkgreen} $\uparrow$ 4.62  & 63.89 \color{darkgreen} $\uparrow$ 5.55  & 80.16 \color{darkgreen} $\uparrow$ 5.01  & 12.95 \color{darkgreen} $\uparrow$ 3.36  & 34.80 \color{darkgreen} $\uparrow$ 4.56  & 26.07 \color{darkgreen} $\uparrow$ 4.30  & 27.50 \color{darkgreen} $\uparrow$ 4.10 \\

Query\textsubscript{reg} $\circ$  Instr\textsubscript{reg}  & 38.57 & 60.47 \color{darkgreen} $\uparrow$ 2.13  & 77.12 \color{darkgreen} $\uparrow$ 1.97  & 10.72 \color{darkgreen} $\uparrow$ 1.13  & 31.72 \color{darkgreen} $\uparrow$ 1.48  & 23.15 \color{darkgreen} $\uparrow$ 1.38  & 24.69 \color{darkgreen} $\uparrow$ 1.29 \\

\bottomrule
\end{tabular}
\caption{ArguAna Full Experimental Results}
\label{tab: ArguAna full experiment result}
\end{table*}

\begin{table*}[!ht]
\centering
\scriptsize
\begin{tabular}{llllllll}
\toprule
Model (Method) & R@5 & R@10 & R@20 & RP & NDCG@10 & MRR@10 & MAP \\
\midrule
\multicolumn{8}{c}{E5-Large-v2} \\
\midrule

Pretrained  & 18.41 & 22.21 & 27.06 & 9.34 & 15.25 & 13.07 & 13.84\\

Contrastive Fine-tuning  & 37.25 & 43.79 & 49.85 & 24.29 & 33.16 & 29.87 & 30.74\\
\cmidrule(lr){1-1}

Promptagator  & 41.53 & 48.34 & 54.66 & 27.23 & 37.07 & 33.56 & 34.43\\
\cmidrule(lr){1-1}

Document\textsubscript{40\% reg}  & 41.40 & 47.56 \color{red} $\downarrow$ -0.78  & 53.69 \color{red} $\downarrow$ -0.97  & 27.83 \color{darkgreen} $\uparrow$ 0.60  & 37.01 & 33.72 & 34.56\\

Query\textsubscript{reg} $\circ$  Document\textsubscript{40\% reg}  & 45.50 \color{darkgreen} $\uparrow$ 3.97  & 51.65 \color{darkgreen} $\uparrow$ 3.31  & 58.08 \color{darkgreen} $\uparrow$ 3.42  & 30.58 \color{darkgreen} $\uparrow$ 3.35  & 40.50 \color{darkgreen} $\uparrow$ 3.43  & 37.01 \color{darkgreen} $\uparrow$ 3.45  & 37.86 \color{darkgreen} $\uparrow$ 3.43 \\

Document\textsubscript{60\% reg}  & 44.15 \color{darkgreen} $\uparrow$ 2.62  & 50.84 \color{darkgreen} $\uparrow$ 2.50  & 55.69 \color{darkgreen} $\uparrow$ 1.03  & 30.07 \color{darkgreen} $\uparrow$ 2.84  & 39.64 \color{darkgreen} $\uparrow$ 2.57  & 36.16 \color{darkgreen} $\uparrow$ 2.60  & 36.93 \color{darkgreen} $\uparrow$ 2.50 \\

Query\textsubscript{reg} $\circ$  Document\textsubscript{60\% reg}  & \textbf{46.23} \color{darkgreen} $\uparrow$ 4.70  & \textbf{52.61} \color{darkgreen} $\uparrow$ 4.27  & \textbf{59.76} \color{darkgreen} $\uparrow$ 5.10  & \textbf{31.75} \color{darkgreen} $\uparrow$ 4.52  & \textbf{41.58} \color{darkgreen} $\uparrow$ 4.51  & \textbf{38.13} \color{darkgreen} $\uparrow$ 4.57  & \textbf{39.02} \color{darkgreen} $\uparrow$ 4.59 \\

Document\textsubscript{80\% reg}  & 28.30 \color{red} $\downarrow$ -13.23  & 33.42 \color{red} $\downarrow$ -14.92  & 39.85 \color{red} $\downarrow$ -14.81  & 16.04 \color{red} $\downarrow$ -11.19  & 24.05 \color{red} $\downarrow$ -13.02  & 21.13 \color{red} $\downarrow$ -12.43  & 22.07 \color{red} $\downarrow$ -12.36 \\

Query\textsubscript{reg} $\circ$  Document\textsubscript{80\% reg}  & 44.80 \color{darkgreen} $\uparrow$ 3.27  & 51.04 \color{darkgreen} $\uparrow$ 2.70  & 58.21 \color{darkgreen} $\uparrow$ 3.55  & 30.87 \color{darkgreen} $\uparrow$ 3.64  & 40.29 \color{darkgreen} $\uparrow$ 3.22  & 36.94 \color{darkgreen} $\uparrow$ 3.38  & 37.85 \color{darkgreen} $\uparrow$ 3.42 \\

Instr\textsubscript{reg}  & 42.15 \color{darkgreen} $\uparrow$ 0.62  & 48.33 & 53.02 \color{red} $\downarrow$ -1.64  & 27.57 & 37.33 & 33.89 \color{darkgreen} $\uparrow$ 0.33  & 34.67\\

Query\textsubscript{reg} $\circ$  Instr\textsubscript{reg}  & 44.90 \color{darkgreen} $\uparrow$ 3.37  & 50.38 \color{darkgreen} $\uparrow$ 2.04  & 56.28 \color{darkgreen} $\uparrow$ 1.62  & 29.07 \color{darkgreen} $\uparrow$ 1.84  & 39.29 \color{darkgreen} $\uparrow$ 2.22  & 35.79 \color{darkgreen} $\uparrow$ 2.23  & 36.66 \color{darkgreen} $\uparrow$ 2.23 \\
\midrule
\multicolumn{8}{c}{SimCSE} \\
\midrule

Pretrained  & 16.96 & 20.76 & 24.98 & 9.62 & 14.58 & 12.67 & 13.44\\

Contrastive Fine-tuning  & 26.78 & 31.74 & 37.15 & 15.73 & 23.11 & 20.42 & 21.28\\
\cmidrule(lr){1-1}

Promptagator  & 30.97 & 37.35 & 44.26 & 18.57 & 27.15 & 23.99 & 24.95\\
\cmidrule(lr){1-1}

Document\textsubscript{40\% reg}  & 32.10 \color{darkgreen} $\uparrow$ 1.13  & 37.43 & 43.07 \color{red} $\downarrow$ -1.19  & 20.29 \color{darkgreen} $\uparrow$ 1.72  & 28.30 \color{darkgreen} $\uparrow$ 1.15  & 25.45 \color{darkgreen} $\uparrow$ 1.46  & 26.34 \color{darkgreen} $\uparrow$ 1.39 \\

Query\textsubscript{reg} $\circ$  Document\textsubscript{40\% reg}  & 34.63 \color{darkgreen} $\uparrow$ 3.66  & 40.71 \color{darkgreen} $\uparrow$ 3.36  & 47.44 \color{darkgreen} $\uparrow$ 3.18  & 22.03 \color{darkgreen} $\uparrow$ 3.46  & 30.64 \color{darkgreen} $\uparrow$ 3.49  & 27.51 \color{darkgreen} $\uparrow$ 3.52  & 28.46 \color{darkgreen} $\uparrow$ 3.51 \\

Document\textsubscript{60\% reg}  & 33.03 \color{darkgreen} $\uparrow$ 2.06  & 38.09 & 43.96 & 20.47 \color{darkgreen} $\uparrow$ 1.90  & 28.87 \color{darkgreen} $\uparrow$ 1.72  & 25.97 \color{darkgreen} $\uparrow$ 1.98  & 26.89 \color{darkgreen} $\uparrow$ 1.94 \\

Query\textsubscript{reg} $\circ$  Document\textsubscript{60\% reg}  & 34.84 \color{darkgreen} $\uparrow$ 3.87  & 40.84 \color{darkgreen} $\uparrow$ 3.49  & 47.08 \color{darkgreen} $\uparrow$ 2.82  & 22.16 \color{darkgreen} $\uparrow$ 3.59  & 30.77 \color{darkgreen} $\uparrow$ 3.62  & 27.64 \color{darkgreen} $\uparrow$ 3.65  & 28.58 \color{darkgreen} $\uparrow$ 3.63 \\

Document\textsubscript{80\% reg}  & 19.75 \color{red} $\downarrow$ -11.22  & 24.46 \color{red} $\downarrow$ -12.89  & 29.89 \color{red} $\downarrow$ -14.37  & 10.20 \color{red} $\downarrow$ -8.37  & 16.73 \color{red} $\downarrow$ -10.42  & 14.33 \color{red} $\downarrow$ -9.66  & 15.20 \color{red} $\downarrow$ -9.75 \\

Query\textsubscript{reg} $\circ$  Document\textsubscript{80\% reg}  & 33.76 \color{darkgreen} $\uparrow$ 2.79  & 39.17 \color{darkgreen} $\uparrow$ 1.82  & 45.63 \color{darkgreen} $\uparrow$ 1.37  & 20.71 \color{darkgreen} $\uparrow$ 2.14  & 29.34 \color{darkgreen} $\uparrow$ 2.19  & 26.26 \color{darkgreen} $\uparrow$ 2.27  & 27.21 \color{darkgreen} $\uparrow$ 2.26 \\

Instr\textsubscript{reg}  & 33.81 \color{darkgreen} $\uparrow$ 2.84  & 39.23 \color{darkgreen} $\uparrow$ 1.88  & 45.79 \color{darkgreen} $\uparrow$ 1.53  & 21.61 \color{darkgreen} $\uparrow$ 3.04  & 29.74 \color{darkgreen} $\uparrow$ 2.59  & 26.79 \color{darkgreen} $\uparrow$ 2.80  & 27.72 \color{darkgreen} $\uparrow$ 2.77 \\

Query\textsubscript{reg} $\circ$  Instr\textsubscript{reg}  & \textbf{37.34} \color{darkgreen} $\uparrow$ 6.37  & \textbf{43.37} \color{darkgreen} $\uparrow$ 6.02  & \textbf{50.45} \color{darkgreen} $\uparrow$ 6.19  & \textbf{24.57} \color{darkgreen} $\uparrow$ 6.00  & \textbf{33.18} \color{darkgreen} $\uparrow$ 6.03  & \textbf{30.02} \color{darkgreen} $\uparrow$ 6.03  & \textbf{30.99} \color{darkgreen} $\uparrow$ 6.04 \\
\midrule
\multicolumn{8}{c}{RoBERTa} \\
\midrule

Pretrained  & 2.84 & 3.60 & 5.33 & 1.11 & 2.17 & 1.73 & 2.10\\

Contrastive Fine-tuning  & 23.59 & 29.09 & 35.27 & 13.80 & 20.69 & 18.10 & 19.05\\
\cmidrule(lr){1-1}

Promptagator  & 27.23 & 33.07 & 39.14 & 15.83 & 23.69 & 20.78 & 21.74\\
\cmidrule(lr){1-1}

Document\textsubscript{40\% reg}  & 28.30 \color{darkgreen} $\uparrow$ 1.07  & 34.47 \color{darkgreen} $\uparrow$ 1.40  & 40.89 \color{darkgreen} $\uparrow$ 1.75  & 16.94 \color{darkgreen} $\uparrow$ 1.11  & 24.79 \color{darkgreen} $\uparrow$ 1.10  & 21.81 \color{darkgreen} $\uparrow$ 1.03  & 22.77 \color{darkgreen} $\uparrow$ 1.03 \\

Query\textsubscript{reg} $\circ$  Document\textsubscript{40\% reg}  & 32.83 \color{darkgreen} $\uparrow$ 5.60  & 39.55 \color{darkgreen} $\uparrow$ 6.48  & 46.38 \color{darkgreen} $\uparrow$ 7.24  & 20.55 \color{darkgreen} $\uparrow$ 4.72  & 29.05 \color{darkgreen} $\uparrow$ 5.36  & 25.82 \color{darkgreen} $\uparrow$ 5.04  & 26.79 \color{darkgreen} $\uparrow$ 5.05 \\

Document\textsubscript{60\% reg}  & 28.67 \color{darkgreen} $\uparrow$ 1.44  & 35.14 \color{darkgreen} $\uparrow$ 2.07  & 41.49 \color{darkgreen} $\uparrow$ 2.35  & 17.00 \color{darkgreen} $\uparrow$ 1.17  & 25.14 \color{darkgreen} $\uparrow$ 1.45  & 22.06 \color{darkgreen} $\uparrow$ 1.28  & 23.03 \color{darkgreen} $\uparrow$ 1.29 \\

Query\textsubscript{reg} $\circ$  Document\textsubscript{60\% reg}  & 34.30 \color{darkgreen} $\uparrow$ 7.07  & \textbf{40.86} \color{darkgreen} $\uparrow$ 7.79  & \textbf{47.52} \color{darkgreen} $\uparrow$ 8.38  & \textbf{21.29} \color{darkgreen} $\uparrow$ 5.46  & \textbf{30.15} \color{darkgreen} $\uparrow$ 6.46  & \textbf{26.83} \color{darkgreen} $\uparrow$ 6.05  & \textbf{27.82} \color{darkgreen} $\uparrow$ 6.08 \\

Document\textsubscript{80\% reg}  & 21.74 \color{red} $\downarrow$ -5.49  & 27.23 \color{red} $\downarrow$ -5.84  & 33.16 \color{red} $\downarrow$ -5.98  & 12.11 \color{red} $\downarrow$ -3.72  & 18.91 \color{red} $\downarrow$ -4.78  & 16.35 \color{red} $\downarrow$ -4.43  & 17.23 \color{red} $\downarrow$ -4.51 \\

Query\textsubscript{reg} $\circ$  Document\textsubscript{80\% reg}  & \textbf{33.69} \color{darkgreen} $\uparrow$ 6.46  & 39.74 \color{darkgreen} $\uparrow$ 6.67  & 46.23 \color{darkgreen} $\uparrow$ 7.09  & 20.69 \color{darkgreen} $\uparrow$ 4.86  & 29.41 \color{darkgreen} $\uparrow$ 5.72  & 26.20 \color{darkgreen} $\uparrow$ 5.42  & 27.18 \color{darkgreen} $\uparrow$ 5.44 \\

Instr\textsubscript{reg}  & 28.12 \color{darkgreen} $\uparrow$ 0.89  & 34.36 \color{darkgreen} $\uparrow$ 1.29  & 40.80 \color{darkgreen} $\uparrow$ 1.66  & 16.28 \color{darkgreen} $\uparrow$ 0.45  & 24.51 \color{darkgreen} $\uparrow$ 0.82  & 21.46 \color{darkgreen} $\uparrow$ 0.68  & 22.43 \color{darkgreen} $\uparrow$ 0.69 \\

Query\textsubscript{reg} $\circ$  Instr\textsubscript{reg}  & 33.09 \color{darkgreen} $\uparrow$ 5.86  & 39.65 \color{darkgreen} $\uparrow$ 6.58  & 46.09 \color{darkgreen} $\uparrow$ 6.95  & 20.57 \color{darkgreen} $\uparrow$ 4.74  & 29.28 \color{darkgreen} $\uparrow$ 5.59  & 26.06 \color{darkgreen} $\uparrow$ 5.28  & 27.03 \color{darkgreen} $\uparrow$ 5.29 \\
\midrule
\multicolumn{8}{c}{SPECTER-v2} \\
\midrule

Pretrained  & 2.98 & 4.22 & 5.26 & 0.97 & 2.43 & 1.87 & 2.18\\

Contrastive Fine-tuning  & 7.85 & 9.60 & 12.56 & 3.72 & 6.36 & 5.35 & 5.89\\
\cmidrule(lr){1-1}

Promptagator  & 6.56 & 8.46 & 11.81 & 3.60 & 5.72 & 4.88 & 5.40\\
\cmidrule(lr){1-1}

Document\textsubscript{40\% reg}  & 9.98 \color{darkgreen} $\uparrow$ 3.42  & 12.52 \color{darkgreen} $\uparrow$ 4.06  & 16.12 \color{darkgreen} $\uparrow$ 4.31  & 4.99 \color{darkgreen} $\uparrow$ 1.39  & 8.55 \color{darkgreen} $\uparrow$ 2.83  & 7.32 \color{darkgreen} $\uparrow$ 2.44  & 7.92 \color{darkgreen} $\uparrow$ 2.52 \\

Query\textsubscript{reg} $\circ$  Document\textsubscript{40\% reg}  & 10.84 \color{darkgreen} $\uparrow$ 4.28  & 12.76 \color{darkgreen} $\uparrow$ 4.30  & 16.26 \color{darkgreen} $\uparrow$ 4.45  & 5.30 \color{darkgreen} $\uparrow$ 1.70  & 8.81 \color{darkgreen} $\uparrow$ 3.09  & 7.57 \color{darkgreen} $\uparrow$ 2.69  & 8.18 \color{darkgreen} $\uparrow$ 2.78 \\

Document\textsubscript{60\% reg}  & 8.64 \color{darkgreen} $\uparrow$ 2.08  & 11.42 \color{darkgreen} $\uparrow$ 2.96  & 14.49 \color{darkgreen} $\uparrow$ 2.68  & 4.51 \color{darkgreen} $\uparrow$ 0.91  & 7.46 \color{darkgreen} $\uparrow$ 1.74  & 6.26 \color{darkgreen} $\uparrow$ 1.38  & 6.80 \color{darkgreen} $\uparrow$ 1.40 \\

Query\textsubscript{reg} $\circ$  Document\textsubscript{60\% reg} & 7.54 \color{darkgreen} $\uparrow$ 0.98  & 10.09 \color{darkgreen} $\uparrow$ 1.63  & 13.29 \color{darkgreen} $\uparrow$ 1.48  & 4.03 \color{darkgreen} $\uparrow$ 0.43  & 6.55 \color{darkgreen} $\uparrow$ 0.83  & 5.48 \color{darkgreen} $\uparrow$ 0.60  & 6.01 \color{darkgreen} $\uparrow$ 0.61  \\

Document\textsubscript{80\% reg}  & 7.39 \color{darkgreen} $\uparrow$ 0.83  & 10.27 \color{darkgreen} $\uparrow$ 1.81  & 13.19 \color{darkgreen} $\uparrow$ 1.38  & 3.62 & 6.46 \color{darkgreen} $\uparrow$ 0.74  & 5.31 \color{darkgreen} $\uparrow$ 0.43  & 5.87 \color{darkgreen} $\uparrow$ 0.47 \\

Query\textsubscript{reg} $\circ$  Document\textsubscript{80\% reg}  & 8.14 \color{darkgreen} $\uparrow$ 1.58  & 11.28 \color{darkgreen} $\uparrow$ 2.82  & 14.69 \color{darkgreen} $\uparrow$ 2.88  & 4.43 \color{darkgreen} $\uparrow$ 0.83  & 7.33 \color{darkgreen} $\uparrow$ 1.61  & 6.13 \color{darkgreen} $\uparrow$ 1.25  & 6.69 \color{darkgreen} $\uparrow$ 1.29 \\

Instr\textsubscript{reg}  & 10.49 \color{darkgreen} $\uparrow$ 3.93  & 13.30 \color{darkgreen} $\uparrow$ 4.84  & 16.80 \color{darkgreen} $\uparrow$ 4.99  & 6.05 \color{darkgreen} $\uparrow$ 2.45  & 9.30 \color{darkgreen} $\uparrow$ 3.58  & 8.07 \color{darkgreen} $\uparrow$ 3.19  & 8.67 \color{darkgreen} $\uparrow$ 3.27 \\

Query\textsubscript{reg} $\circ$  Instr\textsubscript{reg}  & \textbf{11.08} \color{darkgreen} $\uparrow$ 4.52  & \textbf{13.83} \color{darkgreen} $\uparrow$ 5.37  & \textbf{17.36} \color{darkgreen} $\uparrow$ 5.55  & \textbf{6.09} \color{darkgreen} $\uparrow$ 2.49  & \textbf{9.59} \color{darkgreen} $\uparrow$ 3.87  & \textbf{8.28} \color{darkgreen} $\uparrow$ 3.40  & \textbf{8.91} \color{darkgreen} $\uparrow$ 3.51 \\

\bottomrule
\end{tabular}
\caption{WhatsThatBook Full Experimental Results}
\label{tab: wtb full experiment result}
\end{table*}

\end{document}